\documentclass{article}

\usepackage{graphicx}

\usepackage[utf8]{inputenc}
\usepackage[T1]{fontenc}
\usepackage{amsmath}

\usepackage{subcaption}
\usepackage{tikz}
\usepackage{tikzscale}
\usetikzlibrary{decorations.pathmorphing}
\DeclareMathOperator\erf{erf}
\usepackage{a4wide}
\usepackage{natbib}
\usepackage{amssymb}

\title{Landau Damping for Non-Maxwellian Distribution Functions}

\author{Riccardo Stucchi$^1$\footnote{riccardo.stucchi@ipp.mpg.de} \and Philipp Lauber$^2$}
\date{
    \footnotesize{$^{1,2}$Technical University of Munich, TUM School of Natural Sciences, Physics Department, James-Franck-Str. 1, 85747 Garching, Germany}\\
    \footnotesize{$^2$Max Planck Institute for Plasma Physics, Boltzmannstr. 2, 85748 Garching, Germany}\\[2ex]
    \normalsize{\today}
}

\begin{document}

\maketitle

\begin{abstract}
Landau damping is one of the cornerstones of plasma physics. In the context of the mathematical framework developed by Landau in his original derivation of Landau damping, we examine the solutions of the linear Vlasov-Poisson system for different equilibrium velocity distribution functions, such as the Maxwellian distribution, $\kappa$ distributions, and cut-off distributions without and with energy diffusion. Specifically, we focus on the full set of roots that the dispersion relation of the linear Vlasov-Poisson system generally admits, and we wonder if the full structure of solutions might hint at a deeper understanding of the Landau damping phenomenon.
\end{abstract}

\section{Introduction}
Landau damping is the paradigmatic example of kinetic effects in plasma physics. It consists of an interaction mechanism between waves and particles in a collisionless plasma, which converts wave energy into particle kinetic energy. The damping phenomenon was mathematically predicted by Lev Landau in 1946 for one-dimensional electrostatic oscillations, and it was later identified in essentially all other modes of collective oscillations in plasmas. 

In this work, we study Landau damping for the linear Vlasov-Poisson system (LVP) in the context of Landau's original initial value approach \citep{Landau1946}.  Under the assumption of a uniform and static ion background, the linear Vlasov-Poisson system (LVP) describes high-frequency 1D electrostatic oscillations and it is given as 
\begin{equation}\label{eq:VlasovL}
\frac{\partial{f_{1}(x,v,t)}}{\partial{t}} + v \frac{\partial{f_{1}(x,v,t)}}{\partial{x}} - \frac{e}{m} E(x,t) \frac{\partial{f_{0}(v)}}{\partial{v}} = 0,
\end{equation}
\begin{equation}
\frac{\partial{E}(x,t)}{\partial{x}} = -\frac{e n_0}{\epsilon_0}\int \mathrm{d}v f_{1}(x,v,t),
\end{equation}
where $E(x,t)$ is the electric field, $e$ is the elementary charge, $m$ is the electron mass and $n_0$ is the equilibrium ion-electron number density. The function $f(x,v,t)=f_0(v)+f_1(x,v,t)$ corresponds to the electron velocity distribution integrated over $v_y$ and $v_z$, split into an equilibrium part $f_{0}(v)$ and a perturbation part $f_{1}(x,v,t)$.  By following Landau and by assuming that $f_0(v)$ is a Maxwellian distribution function, the evolution of the Fourier transformed electric field, $E(k,t)=\int_{-\infty}^{\infty}E(x,t)e^{-ikx}dx$, is given as a linear combination of initial-value modes,
\begin{equation}\label{eq:E_final_Max}
E(k,t) = \sum_n  A_n e^{-ip_nt},
\end{equation}
where $p_n=\omega_n+i\gamma_n$ are the roots, with $\gamma_n<0$, of the dispersion relation 
\begin{equation}\label{eq:DR_Max}
     k^2 - {\omega^2_p \left[ \int_{-\infty}^{\infty} \frac{\partial{f_{0}(v)}/\partial{v}}{v - p/k} \mathrm{d}v  + 2i \pi\frac{\partial f_0}{\partial v}(p/k) \right]} = 0,
\end{equation}
and $\omega_p=\sqrt{e^2n_0/m_e\epsilon_0}$ is the plasma frequency.

The present work is based on two interesting aspects related to the dispersion relation of \eqref{eq:DR_Max}: the total number of roots $p_n$ and the generic evaluation of $\partial f_0/\partial v$ at the complex value $p/k$.

\begin{figure}
    \centering
    \begin{subfigure}{0.45\textwidth}
        \centering
        \includegraphics[width=\linewidth]{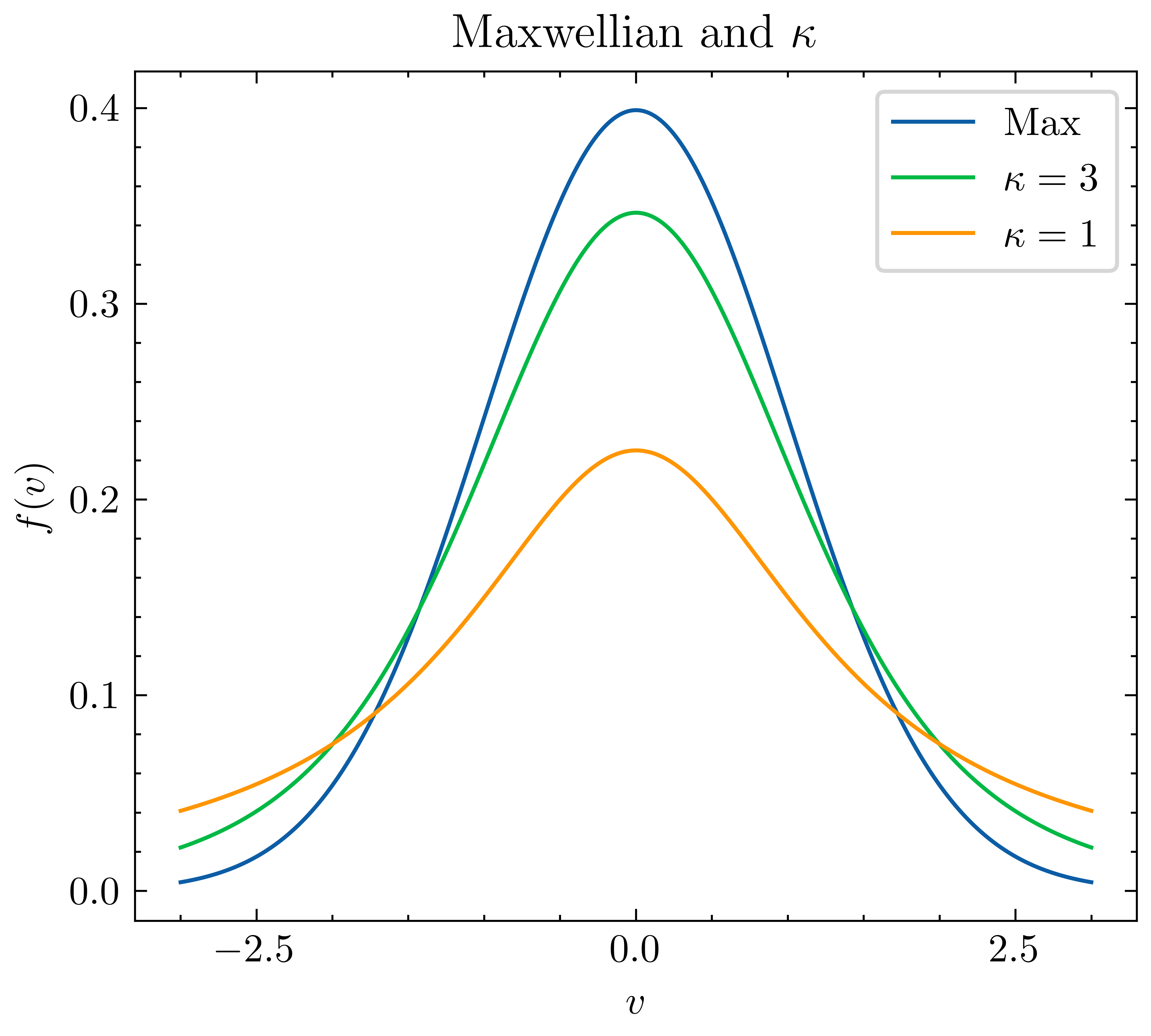} 
    \end{subfigure}
    \hfill
    \hspace{-0.5cm}
    \begin{subfigure}{0.45\textwidth}
        \centering
    \includegraphics[width=\linewidth]{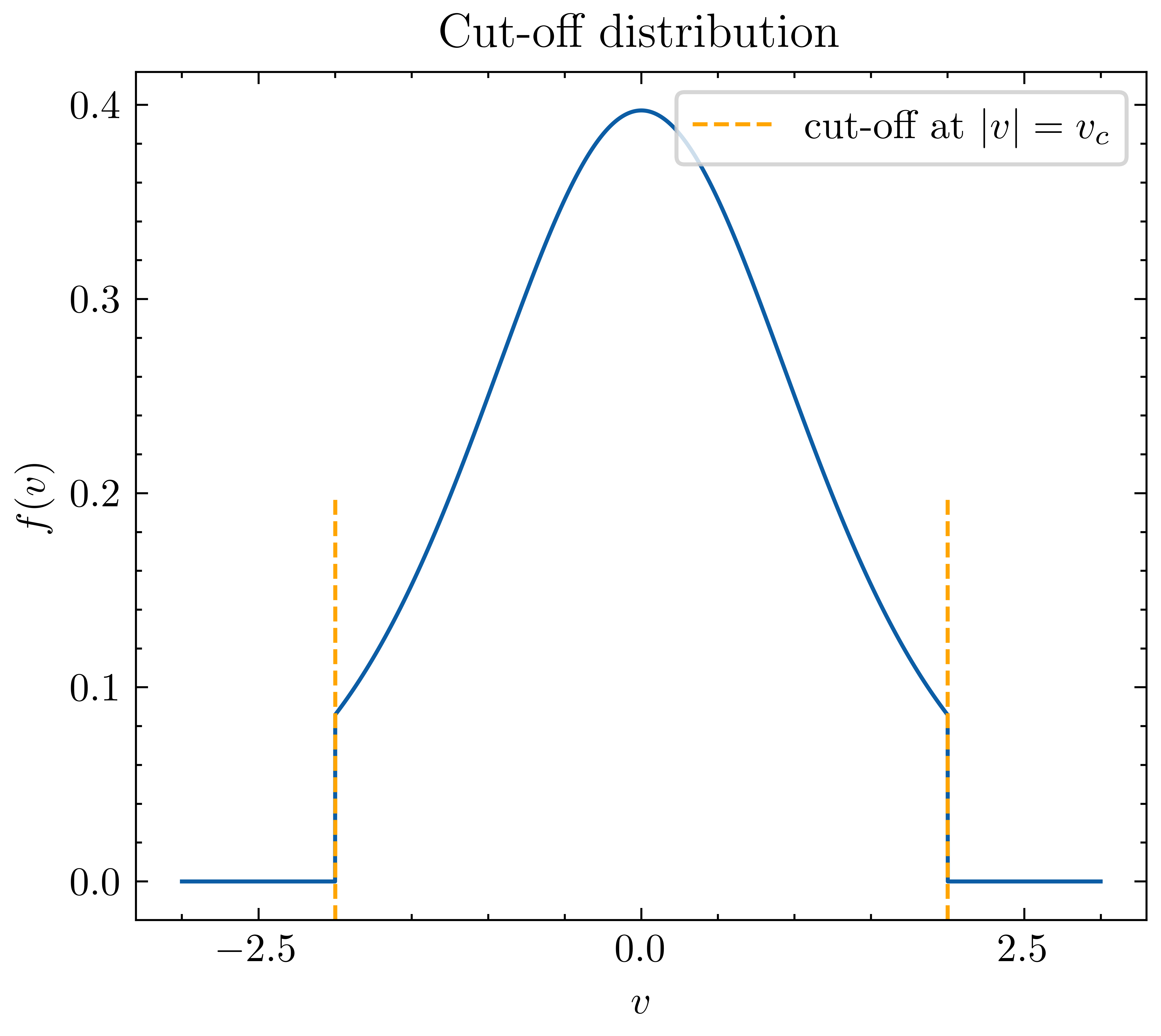} 
    \end{subfigure}
    \vspace{-0.5cm}
    \caption{1D distribution functions. See later sections for analytical definitions.}
    \label{fig:distr}
\end{figure}

Firstly, \eqref{eq:DR_Max} generally admits more than one root. However, in the literature it is common to focus solely on the root $p_n$ with the largest imaginary part $\gamma_n$, as this root dominates the long-time behavior of \eqref{eq:E_final_Max}. The aim of the present
work is then to explore the multiple solutions that LVP admits and investigate how the
structure of solutions is affected by equilibrium distribution
functions $f_0(v)$ that are not Maxwellian, such as the examples of Fig.\,\ref{fig:distr}.

Secondly, velocity distribution functions and their derivatives are generally defined as real-valued functions in the real variable $v\in\mathbb{R}$. Nonetheless, the term $\partial f_0/\partial v(p/k)$ in \eqref{eq:DR_Max}, with $p/k$ complex number, requires the redefinition of $\partial f_0/\partial v$ as a complex variable function. For some distribution functions, e.g., Maxwellian, $\kappa$ distributions (with integer $\kappa$), the redefinition can be performed naturally by simply replacing the real variable $v\in\mathbb{R}$ with a complex one $z\in\mathbb{C}$. For others, e.g. incomplete distributions and $\kappa$ distributions with non-integer $\kappa$, the redefinition can be performed neither continuously nor uniquely. The distinction between two groups of distribution functions, for which the term $\partial f_0/\partial v(p/k)$ is either well or ill-defined,  leads us to organize this work as follows. 

In Sec.\,\ref{Sec2} we follow the literature and provide the necessary mathematical background. In addition, we discuss how the assumption of Maxwellian $f_0(v)$ can be relaxed and how the dispersion relation of \eqref{eq:DR_Max}  can be considered valid for non-Maxwellian distributions. 

In Sec.\,\ref{Sec3} we deal with the first group of distribution functions. Specifically, we show a simple argument to establish the number of roots, we describe analytically the strongly damped roots in the Maxwellian $f_0(v)$ case, and we advance a tentative explanation for the physical meaning of the different solutions of LVP.

In Sec.\,\ref{Sec4} we consider the second class of distribution functions. We mainly focus on the case of distribution functions with compact support, also known as cut-off distributions, that are defined through the Heaviside step function $H(v)$ and that, as an example, are relevant for the description of non-thermal ions in fusion plasmas. We notice that different definitions are possible for the dispersion relation, and although this leads to the same $E$ field evolution, we wonder if such an ambiguity might hide some deeper physical meaning. To investigate this, we employ different sigmoid functions to replace the Heaviside step function and observe how different root structures arise. Lastly, we briefly take two other cases into account, namely the slowing-down distribution function and $\kappa$ distributions with non-integer $\kappa$ parameters, and show the respective pole structures.

As a final comment, it is worth mentioning that the plots of the dispersion relation in the form of \eqref{eq:DR_Max} are the main investigation tool of this paper and, as analytical expressions are not always available, we heavily resort to the numerical scheme from \cite{Xie2013}, which allows the numerical evaluation of the integral in \eqref{eq:DR_Max} for arbitrary distribution functions $f_0(v)$.

\section{The Linear Vlasov-Poisson System:  Solution for $E(k,t)$ and Assumptions on $f_0(v)$}\label{Sec2}
The approach adopted in \cite{Landau1946} consists of applying a Laplace transform in time and a Fourier transform in space to LVP. After some algebra, the Laplace-Fourier transform of the electric field $\tilde{E}_{\textit{UHP}}(p,k)$ can be analytically defined in the upper half-plane (\textit{UHP}, $\mathrm{Im}(p)>0$) as
\begin{equation}
    \tilde{E}_{\textit{UHP}}(k,p) = \frac{G_{\textit{UHP}}(k,p)}{\epsilon_{\textit{UHP}}(k,p)},
\end{equation}
where $G_{\textit{UHP}}$ and $\epsilon_{\textit{UHP}}$ correspond to
\begin{equation}
\begin{split}
    G_{\textit{UHP}}(k,p) &= \frac{e}{k\epsilon_0}\int_{-\infty}^{\infty}\frac{\tilde{f}_{1}(t=0,k,v)}{p-kv}\mathrm{d}v,\,\,\,\,\,\,\,\,\, \mathrm{Im}(p)>0, \\
    \epsilon_{\textit{UHP}}(k,p) &= {1+\frac{e^2}{k\epsilon_0m}\int_{-\infty}^{\infty} \frac{\partial{f_{0}(v)}/\partial{v}}{p-kv} \mathrm{d}v}, \,\,\, \mathrm{Im}(p)>0.
\end{split}
\end{equation}
Here, $\tilde{f}_{1}(t=0,k,v)$ is the Fourier transform of the distribution function perturbation at time $t=0$, $f_1(t=0,x,v)$. 

In its most general form, the solution for the electric field is then given as the inverse Laplace transform of $\tilde{E}_{\textit{UHP}}(k,p)$,
\begin{equation}\label{eq:E(t,x)}
E(k,t) = \frac{1}{2\pi} \int_{i\sigma-\infty}^{i\sigma+\infty} e^{-ipt} \tilde{E}_{\textit{UHP}}(k,p)dp,
\end{equation}
where the $p$-integration is performed on the horizontal line $p=i\sigma + t$, with $t\in [-\infty,\infty]$. $\sigma$ is defined such that $\sigma>\text{Max}_n\left\{0,\mathrm{Im}{(p_n)}\right\}$, i.e., the integration path lies in \textit{UHP} and above all the \textit{UHP} singularities $\{p_n\}_{\textit{UHP}}$ of $\tilde{E}_{\textit{UHP}}(p,k)$. We remark that $\tilde{E}_{\textit{UHP}}(k,p)$ has been defined only in \textit{UHP}, as $\tilde{G}_{\textit{UHP}}(k,p)$ and $\tilde{\epsilon}_{\textit{UHP}}(k,p)$ are discontinuous as $\mathrm{Im}(p)\to0^{\pm}$ and are ill-defined when $\mathrm{Im}(p)=0$.

Although it can already be considered the final solution for the electric field, \eqref{eq:E(t,x)} can be more conveniently recast into the sum of initial value modes by exploiting the residue theorem. The integration contour over $p=i\sigma + t$ is then closed with an infinite radius semicircle in the lower half-plane (\textit{LHP}) and the functions $G_{\textit{UHP}}$ and $\epsilon_{\textit{UHP}}$ are analytically extended to $\mathrm{Im}(p)\leq 0$ by replacing the integral $\int_{-\infty}^{\infty}$ with the Landau integration contour $\int_{LC}$. Specifically, for a generic function $F(z)$, the integral $\int_{-\infty}^{\infty}dvF(v)/(v-z)$ becomes \footnote{the infinite radius semicircle contribution is zero.}
\begin{equation} \label{eq:GPDF}
 Z(z,F)=\int_{LC} {\frac{F(v)}{v-z}\mathrm{d}v} = \begin{cases} \int_{-\infty}^{\infty} \frac{F\left(v\right)}{v-z} \mathrm{d}v, & \mathrm{Im}(z)>0, \\ \mathrm{P.V.}\!\! \int_{-\infty}^{\infty} \frac{F\left(v\right)}{v-z} \mathrm{d}v+i\pi F(z), & \mathrm{Im}(z)=0, \\ \int_{-\infty}^{\infty} \frac{F\left(v\right)}{v-z} \mathrm{d}v+2 i\pi F(z), & \mathrm{Im}(z)<0.\end{cases}
\end{equation}
$G_{\textit{UHP}}$ and $\epsilon_{\textit{UHP}}$ are thus extended as
\begin{equation}
\begin{split}
    G(k,p) &= \frac{e}{k\epsilon_0}\int_{LC}\frac{f_{1}(k,v,t=0)}{p-kv}\mathrm{d}v, \\
    \epsilon(k,p) &= {1+\frac{e^2}{k\epsilon_0m}\int_{LC}\frac{\partial{f_{0}(v)}/\partial{v}}{p-kv} \mathrm{d}v} = 1 - \frac{\omega^2_p}{k^2} Z(p/k,f'_0),
\end{split}
\end{equation}
and $\tilde{E}(k,p)$ is redefined as,
\begin{equation}
    \tilde{E}(k,p) = \frac{G(k,p)}{\epsilon(k,p)}.
\end{equation}

If $f_1(k,v,t=0)$ and $f_0(z)$ are entire functions in the complex variable $z$, $G$ and $\epsilon$ are also entire functions, and the only singularities of $\tilde{E}(p,k)$ are the zeros of $\epsilon(k,p)$. Applying the residue theorem then yields
\begin{equation}\label{eq:E_final}
E(k,t) = \frac{1}{2\pi} \int_{i\sigma-\infty}^{i\sigma+\infty} e^{-ipt} \frac{G(k,p)}{\epsilon(k,p)} dp = - i \sum_n  e^{-ip_nt} \frac{G(k,p_n)}{\partial \epsilon / \partial p|_{p_n}},
\end{equation}
with $p_n$ denoting the roots of the dispersion relation $\epsilon(k,p_n)=0$. Moreover, it can be shown that if $f_0(v)$ has only one maximum, no roots with $\gamma_n\ge0$ are to be expected, and the dispersion relation assumes the form of \eqref{eq:DR_Max} \citep{STIX}.

Let us now comment on Landau's analyticity requirements for ${f}_{1}(k,z,t=0)$ and $f_{0}(z)$ in the complex $z$ variable, namely their analyticity across the whole complex plane (i.e. entire functions). If the final solution is to be expressed as the sum of exponentials with frequencies exclusively given as roots of the dielectric function, the assumption on ${f}_{1}(k,z,t=0)$ cannot be abandoned. Abandoning this assumption introduces $G(p,k)$ singularities that the residue theorem must account for. In the present work, we follow Landau and we always keep ${f}_{1}(k,z,t=0)$ entire. For a discussion on 'non-Landau solutions' induced by disregarding the requirement on ${f}_{1}(k,z,t=0)$, see \citet{belmont2008}. 

On the other hand, \eqref{eq:E_final} remains valid if the assumption on $f_0(z)$ is relaxed by simply requiring it to be meromorphic, i.e., a complex function that is analytical on the whole complex plane except for a set of isolated points. In this case, the singularities of $f_0$ do not pose any issue for the residue theorem because an isolated $\epsilon(p,k)$ singularity translates to an analyticity point of $1/\epsilon(p,k)$. As we show in the next section, singularities of $f_0(z)$ are also meaningful as they can be considered the generators of the roots of the dispersion relation.  Entire and meromorphic functions $f_0(z)$ correspond to the aforementioned first group of distribution functions $f_0(v)$, which admit a straightforward complex redefinition and are dealt with in the next section.

Additionally, distribution functions $f_0(v)$ that cannot be redefined continuously in the whole complex plane are associated with some $f_0(z)$ that is neither entire nor meromorphic. The function $f_0(z)$, and consequently $\tilde{E}(k,p)$, thus feature non-isolated discontinuity points that the residue theorem must avoid.  As shown in \cite{Twiss1952} and \cite{Weitzner1963}, the residue theorem has to circumvent such discontinuities and \eqref{eq:E(t,x)} has to be corrected as
\begin{equation}\label{eq:DR*}
\begin{split}
    E(k,t) &= \frac{1}{2\pi}\int_{i\sigma-\infty}^{i\sigma+\infty} e^{-ipt} \tilde{E}(k,p)\mathrm{d}p \\
    &= -i\sum_n\text{Res}\left\{e^{-ip_nt} \tilde{E}(k,p_n)\right\} -   \frac{1}{2\pi}\int_{\Gamma_D}e^{-ipt} \tilde{E}(k,p)\mathrm{d}p.
\end{split}
\end{equation}
where $\Gamma_D$ indicates the contour around the non-isolated discontinuity of the Laplace transformed electric field, $\tilde{E}(k,p)$, and $p_n$ are the roots of the dispersion relation $\epsilon(k,p_n)=0$.  An interesting feature of \eqref{eq:DR*} is that, despite leading to the same electric field $E(k,t)$, its right hand side can be defined in multiple ways. Sec.\,\ref{Sec4} goes into more details of such a feature.

\section{Distribution Functions With Unique Complex Definition}\label{Sec3}
In this section we consider the case of equilibrium distribution functions $f_0(v)$ that can be uniquely redefined as an entire or meromorphic complex-variable function, $f_0(z)$. Notable examples are the well-known Maxwellian distribution
\[f_{Max}(v) = \frac{1}{\sqrt{\pi v_t^2}}e^{-v^2/v_t^2},\]
and $\kappa$ distributions with integer $\kappa$, of which we consider the 1D version from \cite{Lima2000},
\begin{equation}\label{eq:kappa_def}
    f_{\kappa}(v) = A_{\kappa}\left(1+\frac{v^2}{\kappa v_t^2}\right)^{-\kappa}, 
\end{equation}
where $A_\kappa$ is the normalization constant and $v_t$ is defined as $\sqrt{2T/m}$, with $T$ corresponding to a generic temperature parameter.
While the Maxwellian distribution represents the standard equilibrium state of statistical physics and all plasma physics, $\kappa$ distributions are the non-thermal stationary states of non-extensive statistical mechanics, and it has been shown experimentally that they well describe low collisionality plasma systems, such as the solar wind and the solar corona \citep{Lazar2021}.

\subsection{Number of solutions}
If we assume an entire or meromorphic equilibrium distribution function $f_0(z)$, that admits only one maximum on the real axis, all the roots of the dispersion relation $p_n=\omega_n + i\gamma_n$ lie in the lower half plane (\textit{LHP}) and the dispersion relation $\epsilon(k,p)=0$ is written as \footnote{To improve clarity of notation, we denote the analytical continuation of $\partial f_0(v)/\partial v$ as $f'_0(z)$.}
\begin{equation}\label{eq:DR_LHP}
 k^2 = \omega_p^2 \int_{-\infty}^{\infty}{\frac{f'_0(v)}{v-z}\mathrm{d}v} +2i \pi\omega_p^2 f'_0(z), \quad \mathrm{Im}(z)<0,
\end{equation}
with $z=p/k$.
In the limit $k\to\infty$, the integral term can be neglected and \eqref{eq:DR_LHP} becomes
\begin{equation}\label{eq:DR_LHP_klim}
 \frac{k^2}{\omega_p^2} = Z(z,f'_0(z)) \approx 2 i \pi f'_0(z).
\end{equation}
As $k$ can be arbitrarily large, the previous equality can be satisfied only if its right-hand side can also be made arbitrarily large. In other words, $z$ must belong to a sufficiently small neighborhood of a singularity point of $f'_0(z)$. 
Let us then assume that $z$ belongs to a small neighborhood of $z_0$, a singularity point of order $n$ in \textit{LHP}. By representing $z-z_0$ as $Re^{i\theta}$, we can approximate $f'_0(z)$ as
\begin{equation}\label{eq:Laurent} 
f'_0(z) \approx \frac{C(z_0)e^{i\alpha(z_0)}}{R^n e^{in\theta}},
\end{equation}
where $C(z_0)e^{i\alpha(z_0)}$ is the $n$-order coefficient of the Laurent series expansion. Plugging \eqref{eq:Laurent} into \eqref{eq:DR_LHP_klim} and taking the absolute value and the phase of the resulting equation yield
\begin{equation}\label{eq:kappa_approx}
\begin{cases}
k^2 = \frac{\pi \omega_p^2 C}{R^n}   \rightarrow R_n = \sqrt[n]{\frac{\pi \omega_p^2 C}{k^2}},\\
2m\pi = \frac{\pi}{2} +\alpha -n\theta     \rightarrow \theta_{n,m} = -\frac{2m\pi}{n} + \frac{\pi/2 + \alpha}{n}.
\end{cases}
\end{equation}
Hence, given $n$ (and $k$), the approximate solutions $z_n$ can be expressed as $z_0+R_ne^{i\theta{n,m}}$. 
We then conclude that, as $k\to\infty$, each singularity $z_s$ in \textit{LHP}, of order $n_s$, is generating $n_s$ roots $p_s$ of the dispersion relation of \eqref{eq:DR_LHP}. Moreover, such roots are distributed symmetrically around the singularity $z_0$.
\begin{figure}
    \centering
    \begin{subfigure}{0.48\textwidth}
        \centering
        \includegraphics[width=\linewidth]{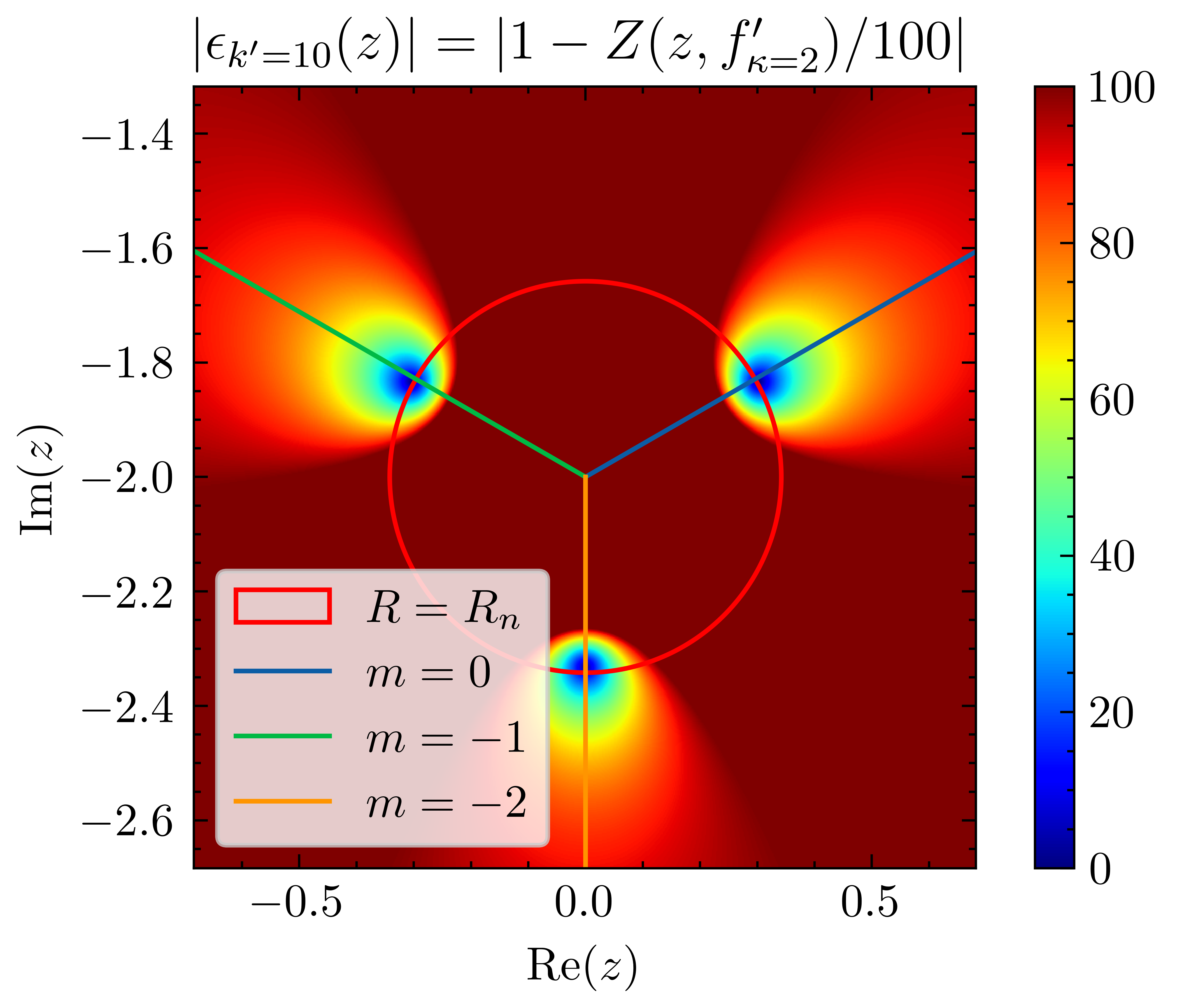} 
     \end{subfigure}
    \hfill
    \begin{subfigure}{0.48\textwidth}
        \centering
    \includegraphics[width=\linewidth]{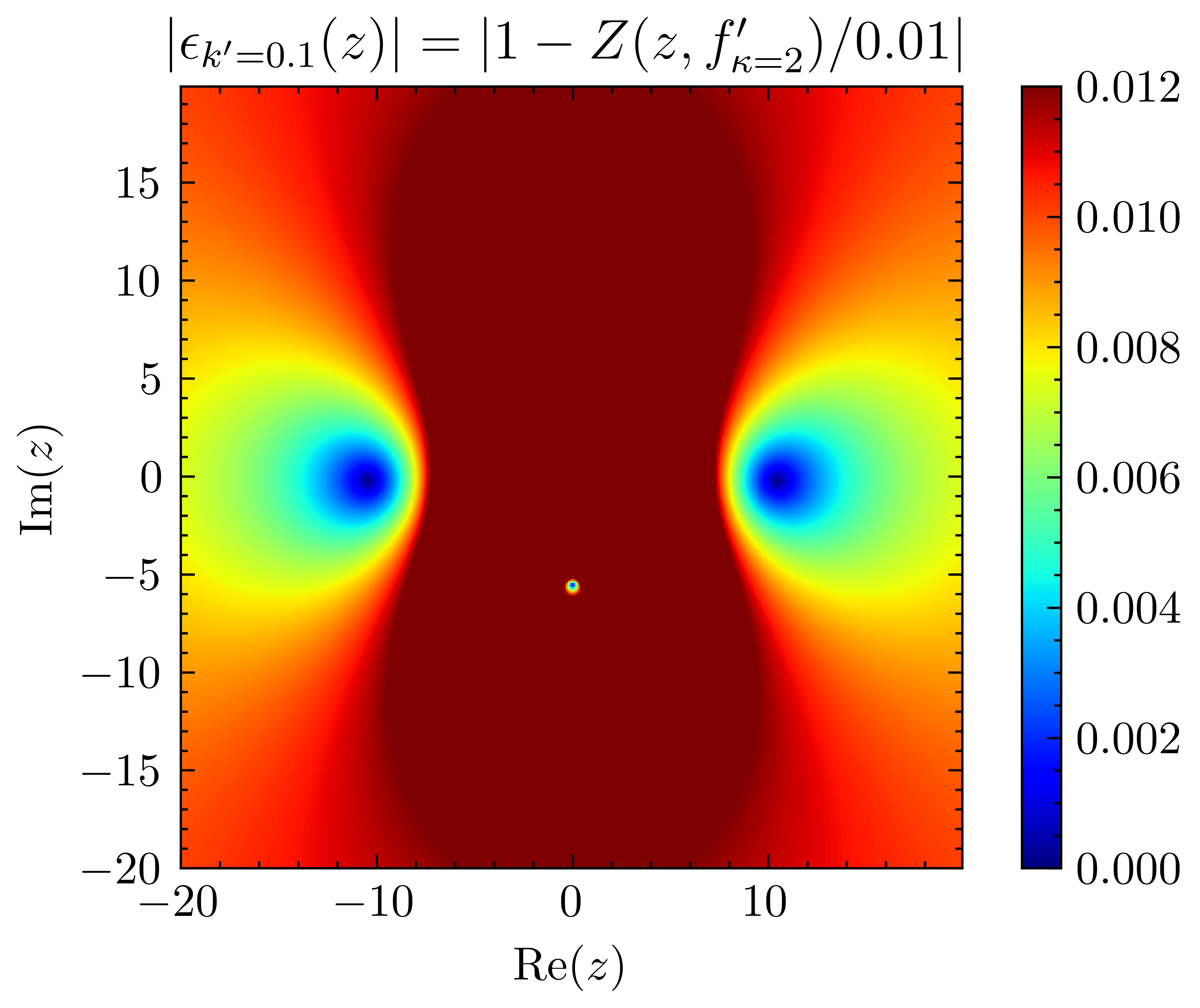} 
    \end{subfigure}
    \vspace{-0.5cm}
    \caption{Dispersion relation roots for the $f_0=f_{\kappa=2}$ case, with $v_t=\sqrt 2$. $k$ is rescaled to the dimensionless $k'=k\lambda_D$, with $\lambda_D=v_t/(\omega_p\sqrt2)$. (\textbf{Left}) $k'=10$ case and comparison with \eqref{eq:kappa_approx}. (\textbf{Right}) $k'=0.1$ case.}
    \label{fig:kappa_approx}
\end{figure}

As an example, Fig.\,\ref{fig:kappa_approx} shows the roots for a $\kappa=2$ distribution $f_{\kappa=2}$ for two different $k$ values. In agreement with our previous considerations, the plot on the left depicts the short-wavelength limit $k\to\infty$. It exhibits 3 roots, as many as the order of the $f'_{\kappa=2}$ singularity, distributed symmetrically around the singularity $z_0=-i\sqrt{\kappa v_t^2}$. Moreover, the intersections
between the red circle and the half-lines $m = i$ give the location of the roots
according to \eqref{eq:kappa_approx}. 

On the right plot, Fig.\,\ref{fig:kappa_approx} shows instead the opposite long wavelength limit $k\to0$. Compared to the left plot, symmetry is lost because of the non-negligible contribution of the integral term, and the roots are here found in regions where $\partial\epsilon/\partial p$ has different values. Recalling that $\partial \epsilon/\partial p$ appears in the linear combination coefficients of \eqref{eq:E_final}, we deduce that the two small damping roots, in regions of smaller $|\partial\epsilon/\partial p|$, dominate the electric field evolution and that the contribution from the third root is suppressed. This behavior is compatible with the $k\to0$ limit of \citet{Landau1946}, for which two stable roots oscillating at the plasma frequency are predicted. 

As a final comment we remark that we demonstrated how many roots are to be expected only in the case $k\to\infty$. However, even if not supported by a rigorous mathematical proof, observing the root structure at different $\kappa$ and $k$ seems to suggest that the number of solutions for distribution functions with a single singularity point in $\textit{LHP}$ does not depend on the wave vector $k$.

\subsection{Maxwellian}
\begin{figure}
\centering
\includegraphics[width=0.5\textwidth]{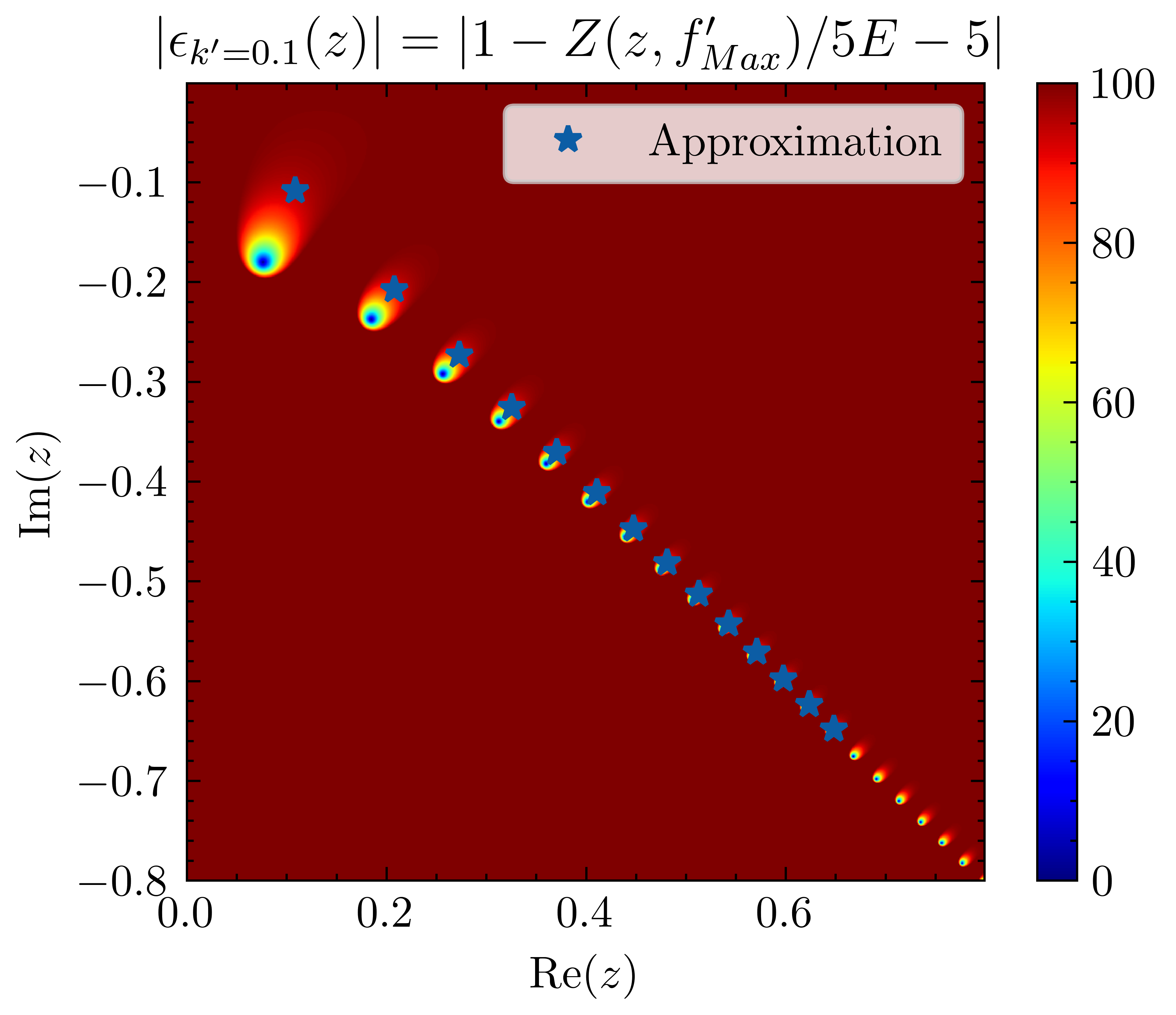}
\caption{Maxwellian multi-roots, solutions of the dispersion relation $k'^2-v_t^2Z_(z,f'_{Max})/2=0$, with thermal velocity $v_t=0.1$, and wave vector $k'=10$. The analytical expression of \eqref{eq:sol_rad} is also represented.}
\label{fig:Max_Multipoles}
\end{figure}

As regards the Maxwellian distribution, we firstly remark that $f_{Max}(v)$ is naturally associated with the entire function $f_{Max}(z)$ by simply replacing $v\in\mathbb{R}$ with $z\in\mathbb{C}$.  In other words, $f_{Max}(z)$ has no singularities at finite real and imaginary parts. Still, it diverges as $\mathrm{Im}(z)$ goes to $-\infty$ so that, to a broader sense, the Maxwellian features a singularity of order infinity at $\mathrm{Im}(z)=-\infty$. Based on the previous discussion about singularities, we then expect infinite solutions to the associated dispersion relation. To describe these solutions analytically,\footnote{The final formula were obtained independently in \citet{maekaku2024time} and \citet{MyThesis2024}.} we neglect the integral term of \eqref{eq:DR_LHP} ($k\to\infty$ limit) and we plug in $f_{Max}(z)$, so that
\begin{equation}
 k^2 \approx 2i\pi \omega_p^2 f'_{Max}(z), 
\end{equation}
and, with $z$ expressed in polar form $Re^{i\theta}$,
\begin{equation}\label{eq:DR_polar_form}
k^2\lambda^2_D  = -2i\sqrt\pi  \frac{Re^{i\theta}}{v_t} \exp\left({-\frac{R^2e^{i\theta}}{v_t^2}}\right).
\end{equation}
Here, we defined $\lambda^2_D$ as the ratio $2v^2_t/\omega^2_p$.
By setting $k'^2=k^2\lambda^2_D$ and by taking the absolute value and the phase of the previous equation, we obtain the system
\begin{equation}\label{eq:abs+ang}
\begin{cases}
k'^2  = 2\sqrt\pi  \frac{R}{v_t} \exp\left({-\frac{R^2\cos{2\theta}}{v_t^2}}\right), \\
2\pi m  = \frac{3\pi}{2} + \theta -  \frac{R^2\sin{2\theta}}{v_t^2}.
\end{cases}
\end{equation}
The first equation of the system in the limit of large $R/v_t$ yields
\begin{equation}\label{eq:delta_app} 
k^2  = 2\sqrt\pi  \frac{R}{v_t} \exp\left({-\frac{R^2\cos{2\theta}}{v_t^2}}\right) \propto \delta(2\theta - \pi/2 - \pi n)
\end{equation}
and implies that $\theta = \frac{\pi}{4} + \frac{\pi}{2} n$. As we are considering only \textit{LHP}, the solutions for $\theta$ are $\theta = \left\{-\frac{\pi}{4},-\frac{3}{4}\pi\right\}$ (with $n=-1,-2$), and by substituting these values into the second equation of \eqref{eq:abs+ang}, we determine
\begin{equation}\label{eq:sol_rad}
       2\pi m + \frac{3\pi}{4} = \frac{R^2}{v_t^2} \to \frac{R}{v_t} =\sqrt{\pm 2\pi m + \frac{3\pi}{4}},
\end{equation}
where the plus and minus sign correspond, respectively, to $\theta = -\frac{\pi}{4}$ and $\theta = -\frac{3}{4}$. It is clear that, due to the assumption of large $R/v_t$ needed for \eqref{eq:delta_app}, the solution for $R/v_t$ becomes more accurate as $m$ grows. The correctness of \eqref{eq:sol_rad}  in the limit $R\to\infty$ is depicted in Fig.\,\ref{fig:Max_Multipoles}.

\subsection{From $\kappa$ to Maxwellian}
\begin{figure}
\centering
\includegraphics[width=1\textwidth]{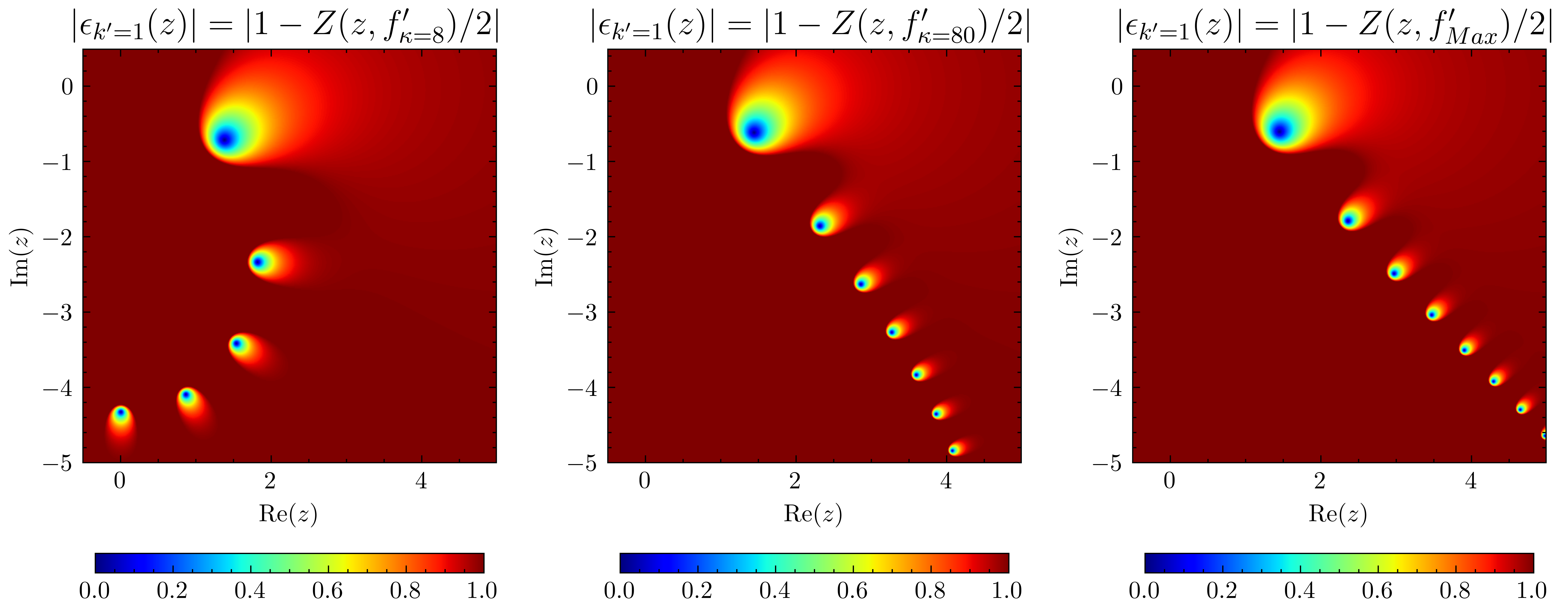}
\caption{Roots for $\kappa$ distributions approaching a Maxwellian. (\textbf{Left}) $f_0=f_{\kappa=8}$. (\textbf{Center}) $f_0=f_{\kappa=80}$. (\textbf{Right}) $f_0=f_{\kappa=\infty}=f_{Max}$. Parameters used: $v_t=\sqrt{2}$, $k'=1$.}
\label{fig:kappa_maxw}
\end{figure}

The typical feature of the 'Maxwellian' LVP is that its dispersion relation admits infinite roots. Mathematically, the Maxwellian root structure can be straightforwardly connected to the root structure for the $\kappa$ distribution case by simply observing the limit
\begin{equation}
f_{\kappa}(z) = A_{\kappa}\left(1+\frac{v^2}{\kappa v_t^2}\right)^{-\kappa} \xrightarrow{\kappa\to\infty}  f_{M} = \frac{1}{\sqrt{\pi}v_t}e^{-v^2/v_t^2}.
\end{equation}
Specifically, as shown in Fig.\,\ref{fig:kappa_maxw}, the Maxwellian root structure can be interpreted as the root structure induced by an equilibrium $\kappa$ distribution with $\kappa$ that goes to infinity and with the singularity $-i\sqrt{\kappa v_t^2}$ that moves towards $\mathrm{Im}(z)=-\infty$. 

If the singularity argument sheds light on how the roots of the LVP dispersion relation arise from a purely mathematical point of view, we wonder if a deeper physical meaning can be associated with the number and the structure of LVP solutions.
A tentative and speculative interpretation is suggested by considering the more precise definition of $\kappa$ distributions in the framework of non-extensive statistical mechanics. In particular, given an $N$-particle system with $f$ degrees of freedom described by an $f$-dimensional $\kappa$ distribution $f_N$, the distribution function for the single degree of freedom is obtained by marginalizing $f_N$ over the $f-1$ degrees of freedom. The more precise version of \eqref{eq:kappa_def} is then given from \citet{liv_2011} as 
\begin{equation}
    f_{\kappa_0}(v) \propto \left(1+\frac{v^2}{\kappa_0 v_t^2}\right)^{-\kappa_0 -3/2}, 
\end{equation}
in terms of the invariant $\kappa_0$ index. The same \citet{liv_2011} proves that $\kappa_0$ represents the thermodynamic distance from the Maxwellian thermal equilibrium and captures the correlation $\rho$ between the degrees of freedom of the system through the expression
$$
    \rho(\kappa_0) = \frac{3D/2}{3D/2 + \kappa_0}, 
$$
with $D$ being the degrees of freedom of the single particle. Hence, as $\kappa_0$ increases, larger collisionality destroys correlation, and in the limit $\kappa_0\to\infty$, the null-correlation Maxwellian distribution is retained. 

Recalling that the number of LVP roots corresponds to the order of the $f'_0(z)$ \textit{LHP} singularity ($\kappa_0+1/2$ for the $f_0=f_{\kappa_0}$ case,\footnote{In accordance to previous comments regarding the definition of the complex variable function $f_0(z)$, we assume that $\kappa_0+1/2$ is an integer. The non-integer $\kappa$ case is discussed later.}) we are consequently tempted to hypothesize a connection between the number of roots and the system correlation so that a stronger (weaker) correlation implies fewer (more) roots and a reduced (increased) 'freedom of movement' of the system. 

\section{Non-unique analytical continuation}\label{Sec4}
For other interesting cases, the complex redefinition $f_0(v)\to f_0(z)$ cannot be performed as straightforwardly as in the previous section. The main notable examples involve distribution functions that feature an empty interval $\mathcal{I}$ such that $f_0(v\in\mathcal{I})\approx0$, e.g.,
\begin{itemize} 
\item Incomplete distributions: in laboratory plasmas, in presence of potential barriers, such as sheats near material walls or probes, a trapped-passing boundary is created. As a consequence, at the edge of an absorbing wall sheath the passing interval of the electron distribution will be empty.
\item Slowing-down distribution: fast ions, such as $3.5$MeV $\alpha$-particles and NBI beams in fusion plasmas, because of their high energy compared to the background and  because of the low intra-species collisionality, do not thermalize to the Maxwellian state. Instead, they slow-down via collisions with the background plasma to a stable distribution known as slowing-down, where  $f_0(|v|>v_c)\approx0$, with $v_c$ standing for the fast species injection velocity \citep{Gaffey1976}. 
\end{itemize}

In both cases, if we neglect the energy dispersion, the empty intervals can be modeled by means of the Heaviside function $H(v)$, which is responsible for the problematic complex redefinition.

A further example, not related to the Heaviside function, corresponds to $\kappa$ distribution with non-integer $\kappa$. In fact, when $\kappa$ is a non-integer, the complex distribution function $f_0(z)$ becomes multivalued and branch cuts are introduced.

In the following, we focus on the case of cut-off distributions, which exemplify the issues connected to the Heaviside function, and we later briefly mention the specific case of slowing-down and non-integer $\kappa$ distributions.

\subsection{Cut-off distributions}
We formally define cut-off distributions as
\begin{equation}
    f_{CO}(v) = 
\begin{cases}
    F_0(v) & \text{if } |v| \leq v_c, \\
    0 & \text{otherwise},
\end{cases}
\end{equation}
and we conveniently express them in terms of the Heaviside function $H(v)$,
\begin{equation}
f_{CO}(v) = H(v+v_c) \cdot F_0(v) \cdot H(-v+v_c) = F_0(v) W_{v_c}(v) .
\end{equation}
Here, we introduce $W_{v_c}(v)$ as the symmetric 'window' function with cut-off velocity $v_c $, representing the product of two Heaviside functions. For simplicity, we further assume that $F_0(v)$ is a symmetric real-valued function with a single maximum and a unique definition in \textit{LHP}\footnote{The case of asymmetric distribution function does not introduce significant differences in the following discussion.}. The derivative $f'_{CO}(v)$ is then given by
\begin{equation}\label{eq:CO_der}
f'_{CO}(v) = F'_0(v) W_{v_c}(v)+F_0(v) \left[\delta(v+v_c) - \delta(v-v_c)\right].
\end{equation}
By considering the definition from \eqref{eq:GPDF} of the 'Landau integral' with $F(v)=f'_{CO}(v)$,
\begin{equation}
Z(z,f'_{CO}) = \int_{LC}\frac{f'_{CO}(v)}{p-kv} \mathrm{d}v,
\end{equation}
we realize that $Z(z,f'_{CO})$ remains consistent and uniquely defined in \textit{UHP} and on the real axis, but not in \textit{LHP}.

\subsubsection{Upper half-plane and real axis}
Given the assumptions on $F_0$, the same arguments from \cite{STIX} can be applied and unstable solutions with $\mathrm{Im}(z)>0$ can be excluded. On the other side, on the real axis the dispersion relation becomes
\begin{equation}\label{eq:DR_R_CS}
\frac{k^2}{\omega_p^2}= \mathrm{P.V.}\!\!\int_{-v_c}^{v_c}{\frac{F'_0(v)}{v-z}\mathrm{d}v} + F_0(v_c) \frac{2v_c}{z^2-v_c^2} + \pi i F'_0(v) W_{v_c}(v), 
\end{equation}
and similarly to \cite{Weitzner1963} we can show that if $F_0(v_c)\neq0$ a stable solution exists for any value of $k$, provided $|\mathrm{Re}(z)|>v_c$.

The real-axis solution remains consistent with the argument that singularities generate solutions, as the singularity point of $f'_{CO}$ at $z=\pm v_c$ is responsible for such a stable root.
In addition, the physical interpretation in terms of the standard resonant interaction mechanism is also straightforward. Compact support implies that no particles at velocities larger than $v_c$ are present. Hence, when the wave phase velocity is greater than $v_c$ there is no resonating particle, and the wave propagates without damping.

Since unstable solutions are absent, the stable solution dominates the long-time limit. Nonetheless, to examine the full-time evolution of $E(k,t)$, the \textit{LHP} must be investigated and $f'_{CO}(z)$ must be defined.

\begin{figure}[t]
\hspace{-0.5cm}
\includegraphics[width=1\textwidth]{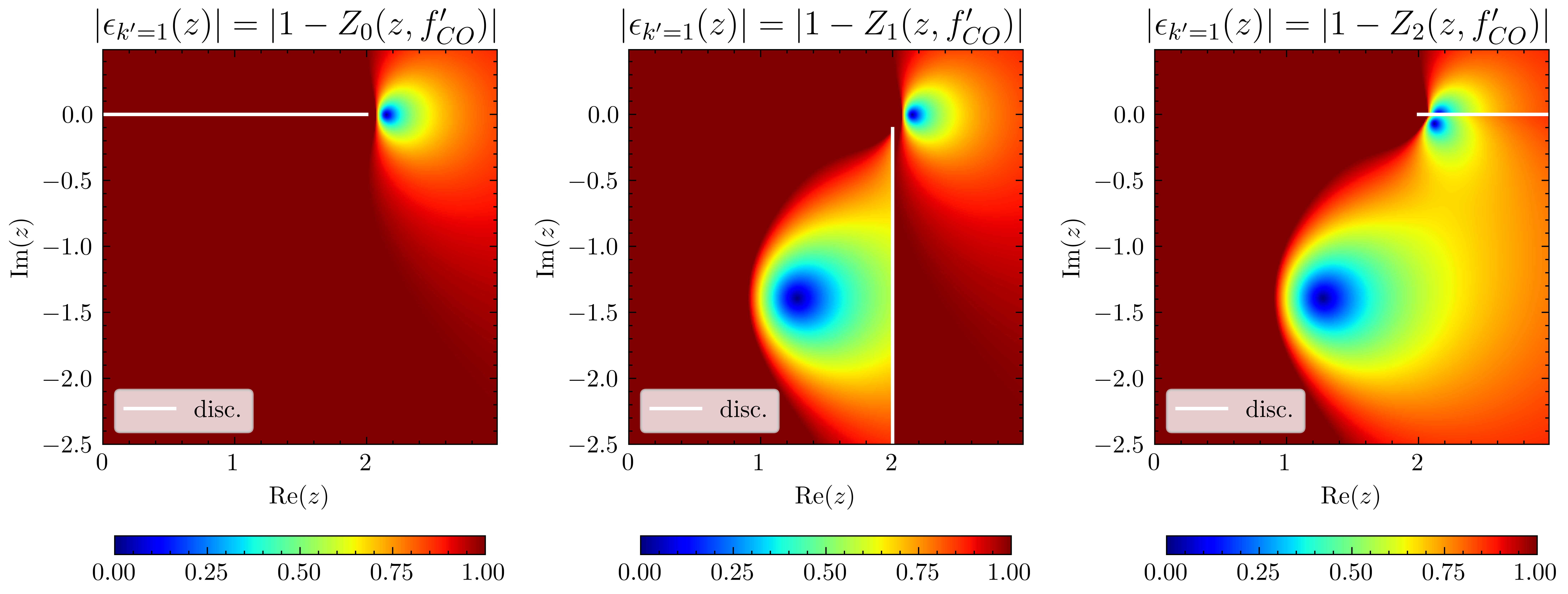}
\caption{Root structures for different \textit{LHP} definitions of $f_{CO}$. The white lines represent the discontinuity of the different $Z_{i}(z)$. The support function $F_0$ is a $\kappa=1$ distribution $f_{\kappa=1}$, with $v_t=\sqrt{2}$}
\label{fig:kappa_co}
\end{figure}

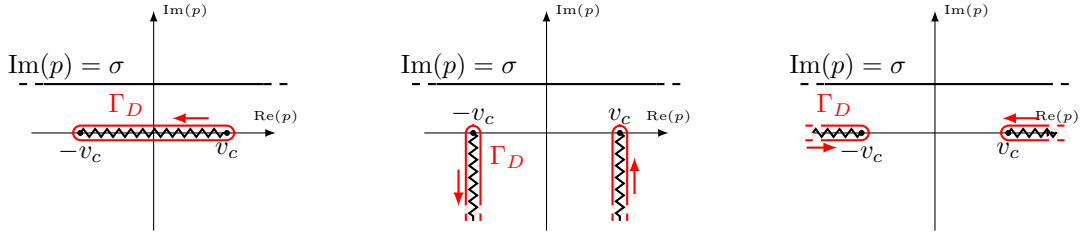
\begin{figure}[t]
  \begin{subfigure}[b]{0.3\textwidth}
    \centering
    \begin{tikzpicture}[scale=0.325]
        \draw[-latex] (-5,0) -- (5,0) node[above, font=\tiny] {$\operatorname{\mathrm{Re}}(p)$};
        \draw[-latex] (0,-4) -- (0,5) node[right, font=\tiny] {$\operatorname{\mathrm{Im}}(p)$};

        \draw[thick] (-4.5,2) -- (4.5,2);
        \draw[dashed,thick] (4.5,2) -- (5.5,2);
        \draw[dashed,thick] (-5.5,2) -- (-4.5,2);

        
        \node[above left] at (-0.8,1.8) {$\mathrm{Im}(p) = \sigma$};
        
        \filldraw (3, 0) circle (0.1) node[below] {$v_c$};
        \filldraw (-3, 0) circle (0.1) node[below] {$-v_c$};
        
        \draw[thick,decorate,decoration={zigzag,segment length=4.5,amplitude=1.5}] (-3, 0) -- (3, 0);
        \draw[red, thick, rounded corners=3pt] (-3.3, -.3) rectangle (3.3, 0.3);
        \node[above left, red] at (0,0.3) {$\Gamma_D$};
        \draw[->, thick, red, >=latex] (2.25,0.6) -- (0.75,0.6); 
        
    \end{tikzpicture}
 
  \end{subfigure}
  \hfill
  \begin{subfigure}[b]{0.3\textwidth}
    \centering
    \begin{tikzpicture}[scale=0.325]
        \draw[-latex] (-5,0) -- (5,0) node[above, font=\tiny] {$\operatorname{\mathrm{Re}}(p)$};
        \draw[-latex] (0,-4) -- (0,5) node[right, font=\tiny] {$\operatorname{\mathrm{Im}}(p)$};

        \draw[thick] (-4.5,2) -- (4.5,2);
        \draw[dashed,thick] (4.5,2) -- (5.5,2);
        \draw[dashed,thick] (-5.5,2) -- (-4.5,2);

        
        \node[above left] at (-0.8,1.8) {$\mathrm{Im}(p) = \sigma$};
        
        \filldraw (3, 0) circle (0.1) node[above] {$v_c$};
        \filldraw (-3, 0) circle (0.1) node[above] {$-v_c$};
        
        \draw[red, thick, rounded corners=3pt] (-3.3, -3.3) rectangle (-2.7, 0.3);
        \draw[white, thick, rounded corners=3pt] (-3.3, -3.3) rectangle (-2.7, -3.1);
        \draw[red, thick, dashed] (-3.3, -3.3) -- (-3.3, -3.6);
        \draw[red, thick, dashed] (-2.7, -3.3) -- (-2.7, -3.6);
        \draw[red, thick, rounded corners=3pt] (3.3, -3.3) rectangle (2.7, 0.3);
        \draw[white, thick, rounded corners=3pt] (3.3, -3.3) rectangle (2.7, -3.1);
        \draw[red, thick, dashed] (3.3, -3.3) -- (3.3, -3.6);
        \draw[red, thick, dashed] (2.7, -3.3) -- (2.7, -3.6);
        \draw[thick,decorate,decoration={zigzag,segment length=4.5,amplitude=1.5}] (-3, 0) -- (-3, -3.6);
        \draw[thick,decorate,decoration={zigzag,segment length=4.5,amplitude=1.5}] (3, 0) -- (3, -3.6);
        \node[right, red] at (-2.7,-1) {$\Gamma_D$};
        \draw[->, thick, red, >=latex] (3.6,-2.5) -- (3.6,-1); 
        \draw[->, thick, red, >=latex] (-3.6,-1.5) -- (-3.6,-3); 
        
    \end{tikzpicture}
 
  \end{subfigure}
  \hfill
  \begin{subfigure}[b]{0.3\textwidth}
    \centering
    \begin{tikzpicture}[scale=0.325]
        \draw[-latex] (-5,0) -- (5,0) node[above, font=\tiny] {$\operatorname{\mathrm{Re}}(p)$};
        \draw[-latex] (0,-4) -- (0,5) node[right, font=\tiny] {$\operatorname{\mathrm{Im}}(p)$};

        \draw[thick] (-4.5,2) -- (4.5,2);
        \draw[dashed,thick] (4.5,2) -- (5.5,2);
        \draw[dashed,thick] (-5.5,2) -- (-4.5,2);

        
        \node[above left] at (-0.8,1.8) {$\mathrm{Im}(p) = \sigma$};
        
        \filldraw (3, 0) circle (0.1) node[below] {$v_c$};
        \filldraw (-3, 0) circle (0.1) node[below] {$-v_c$};
        
        \draw[red, thick, rounded corners=3pt] (2.7, -0.3) rectangle (5, 0.3);
        \draw[white, thick, rounded corners=3pt] (4.7, -0.3) rectangle (5, 0.3);
        \draw[->, thick, red, >=latex] (4.25,0.6) -- (2.75,0.6); 
        \draw[->, thick, red, >=latex] (-5.25,-0.6) -- (-4,-0.6); 
        \draw[red, thick, dashed] (5, 0.3) -- (5.3, 0.3);
        \draw[red, thick, dashed] (5, -0.3) -- (5.3, -0.3);
        \draw[red, thick, rounded corners=3pt] (-2.7, 0.3) rectangle (-5, -0.3);
        \draw[white, thick, rounded corners=3pt] (-4.7, 0.3) rectangle (-5, -0.3);
        \draw[red, thick, dashed] (-5.3, 0.3) -- (-5., 0.3);
        \draw[red, thick, dashed] (-5.3, -0.3) -- (-5., -0.3);
        \draw[thick,decorate,decoration={zigzag,segment length=4.5,amplitude=1.5}] (3, 0) -- (5, 0);
        \draw[thick,decorate,decoration={zigzag,segment length=4.5,amplitude=1.5}] (-5, 0) -- (-3, 0);
        \node[above left, red] at (-3,0.3) {$\Gamma_D$};
    \end{tikzpicture}
  \end{subfigure}
  \caption{Discontinuities generated by the different definitions for the 'window' function, $W_{v_c,0}$, $W_{v_c,1}$, and $W_{v_c,2}$, and the associated contour $\Gamma_D$ needed for \eqref{eq:DR*}.}
  \label{fig:Disc_GammaD}
\end{figure}

\subsubsection{Lower half-plane}
In \textit{LHP}, the formal definition of $Z(z,f')$ for cut-off distributions is
\begin{equation}
Z_{\textit{LHP}}(z,f'_{CO}) = \int_{-\infty}^{\infty}{\frac{F'_0(v) W_{v_c}(v)}{v-z}\mathrm{d}v} + 2\pi i F'_0(z)W_{v_c}(z) + F_0(v_c)\frac{2v_c}{z^2-v_c^2}.
\end{equation}
While $F'_0(z)$ is generally well-defined, $W_{v_c}(z)$ is not. Since there is no unique way to define $W_{v_c}(z)$, we introduce three different possible definitions, among infinitely many, for the 'window' function: 
\begin{equation}\label{eq:H012}
\begin{split}
W_{v_c,0}(z) &= 0,\\
W_{v_c,1}(z) &= H(v_c-|\mathrm{Re}(z)|),\\
W_{v_c,2}(z) &= 1.
\end{split}
\end{equation}
This leads to three different definitions for $Z(z,f'_{CO})$ in \textit{LHP},
\begin{equation}\label{eq:Z012}
\begin{split}
Z_{\textit{LHP},0}(z,f'_{CO}) &= \int_{-v_c}^{v_c}{\frac{F'_0(v)}{v-z}\mathrm{d}v} + F_0(v_c) \frac{2v_c}{z^2-v_c^2},\\
Z_{\textit{LHP},1}(z,f'_0) &= \int_{-v_c}^{v_c}{\frac{F'_{CO}(v)}{v-z}\mathrm{d}v} + 2\pi i F'_0(z)H(v_c-|\mathrm{Re}(z)|) + F_0(v_c) \frac{2v_c}{z^2-v_c^2},\\
Z_{\textit{LHP},2}(z,f'_{CO}) &= \int_{-v_c}^{v_c}{\frac{F'_0(v)}{v-z}\mathrm{d}v} + 2\pi i F'_0(z) + F_0(v_c) \frac{2v_c}{z^2-v_c^2},
\end{split}
\end{equation}
and to three different dispersion relations and associated roots, as depicted in Fig.\;\ref{fig:kappa_co}.

Despite different root structures, \eqref{eq:DR*} guarantees that the electric field evolution does not depend on the specific definition $W_{v_c}(z)$. The key point is to observe that extending the window function $H_d(z)$ into \textit{LHP} leads to non-isolated points of discontinuity for $Z(z,f'_0)$ and $E(p,k)$. When applying the residue theorem, such discontinuities must be circumvented as exemplified in Fig.\,\ref{fig:Disc_GammaD}, and lead to the additional integral term. Nevertheless, we notice that:
\begin{itemize}
\item In the original Landau's treatment, as per \eqref{eq:E_final}, the $E$ field solution could be entirely decomposed into the sum of residues, i.e. initial value modes. On the contrary, when $f'_0(z)$ is not continuous, this is not true as an integral contribution around $\Gamma_D$ remains. Thus,  the whole point of applying the residue theorem to conveniently express the inverse Laplace transform as a sum of residues does no longer hold. We then wonder if the integral contribution $\Gamma_D$ can be expressed, at least approximately, as the sum of initial-value modes.
\item Although the final result is the same, multiple equivalent descriptions are possible in \textit{LHP}. Again, we wonder if the \textit{LHP} extension is a pure mathematical artifact or whether it might hide some physical meaning.
\end{itemize}
To tackle these aspects, we include energy dispersion in the equilibrium distribution function by smoothing the Heaviside function by means of sigmoid functions.

\subsection{Smooth cut-off distributions}
\begin{figure}[h]
\centering
\includegraphics[width=0.5\textwidth]{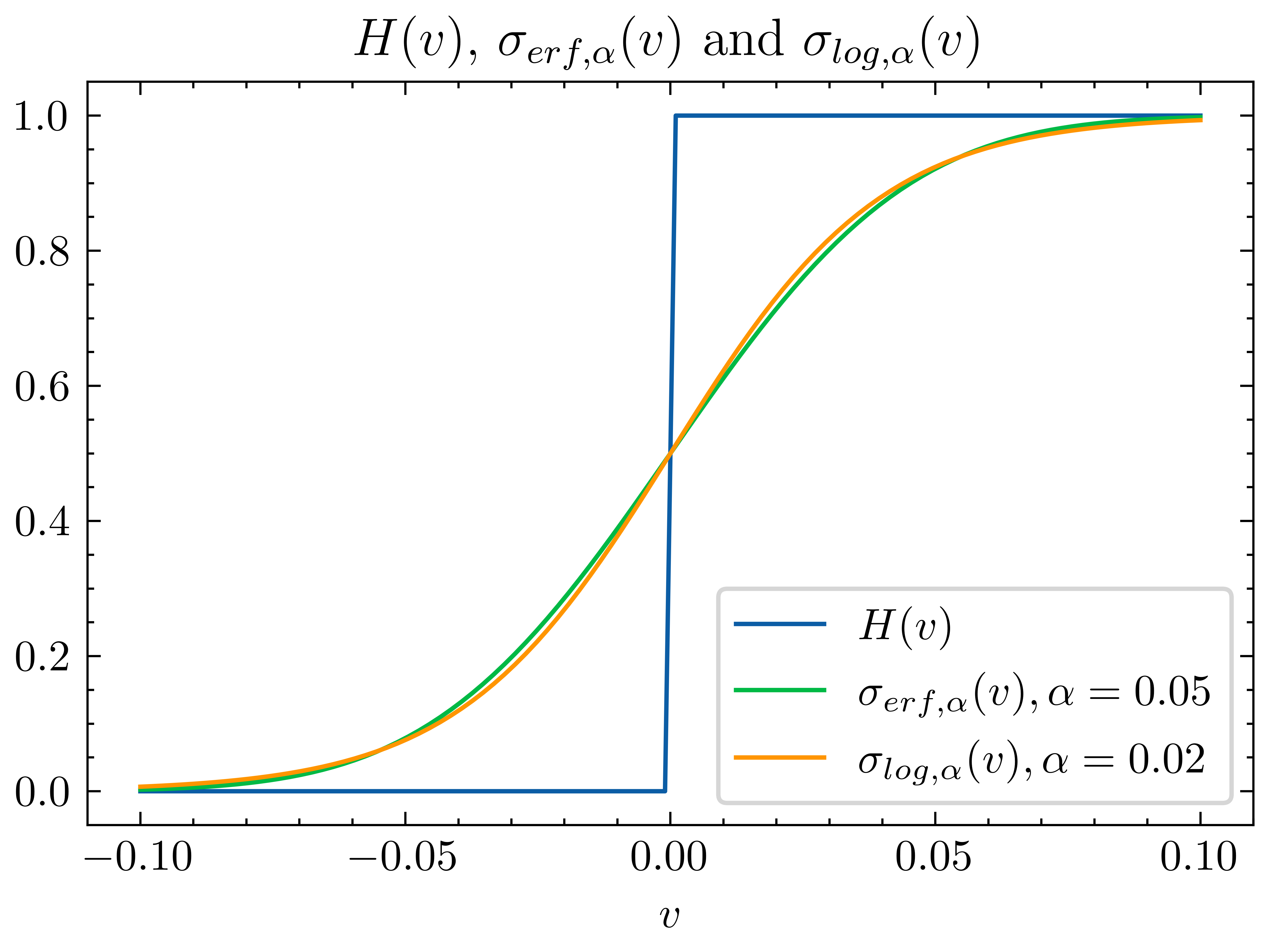}
\caption{Logistic and error function sigmoids. The $\alpha$ parameters are chosen in order to approximately overlap the sigmoids.}
\label{fig:sigmoids}
\end{figure}

\begin{figure}
    \centering
    \begin{subfigure}{0.48\textwidth}
        \centering
        \includegraphics[width=\linewidth]{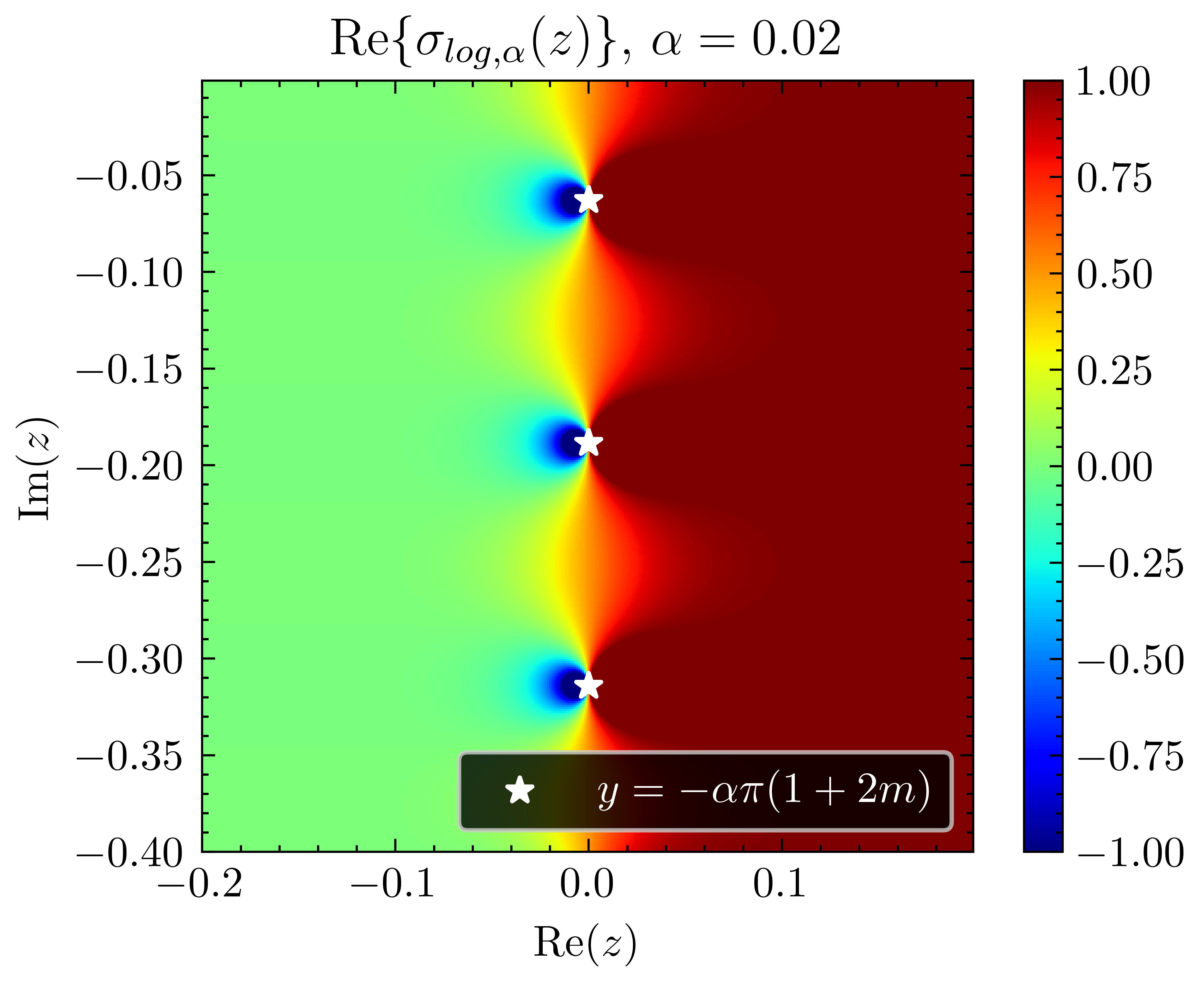} 
      \end{subfigure}
    \hfill
    \begin{subfigure}{0.48\textwidth}
        \centering
    \includegraphics[width=\linewidth]{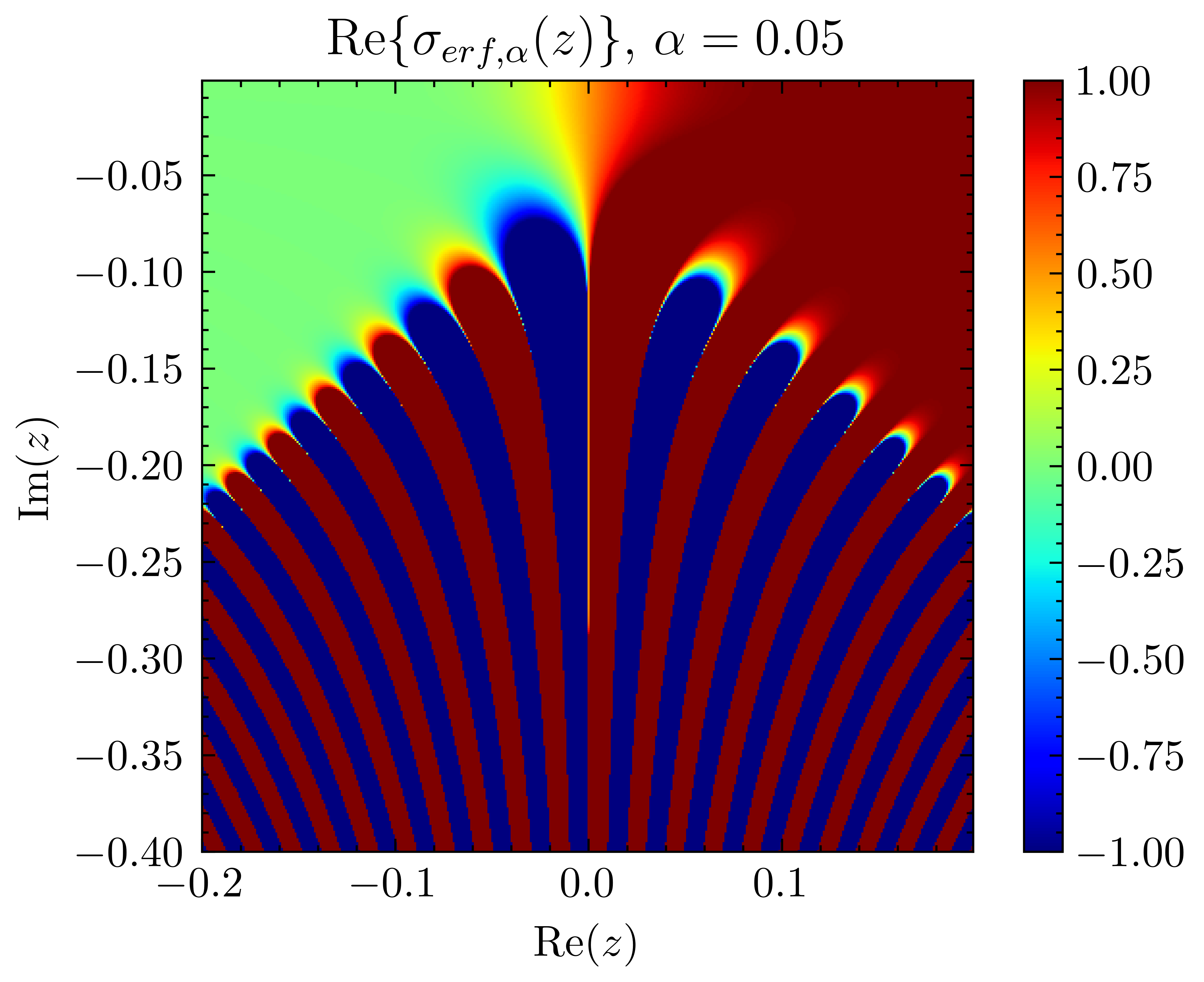} 
    \end{subfigure}
    \vspace{-0.5cm}
    \caption{Real part of the complex variable logistic ($\sigma_{\log,\alpha}(z)$, \textbf{Left}) and error function ($\sigma_{\erf,\alpha}(z)$, \textbf{Right}) sigmoid. The $\alpha$ parameters are chosen so that the two sigmoids approximately overlap on the real axis. The singularities of $\sigma_{\log,\alpha}(z)$ are also shown.}
    \label{fig:erf_log}
\end{figure}

Because of energy dispersion, a cut-off distribution more realistically entails a high-energy tail rather than being sharply cut-off at $v=v_c$. We model such energy dispersion in two possible ways: with a logistic sigmoid, 
\begin{equation}\label{eq:log}
\sigma_{\log,\alpha}(v) = \frac{1}{1+e^{-v/\alpha}} \xrightarrow{\quad \alpha\to0 \quad} H(v) = \begin{cases}1  &v>0, \\ 1/2  &v=0, \\ 0  &v<0, \end{cases}
\end{equation}
and an error function sigmoid,
\begin{equation}
\sigma_{\erf,\alpha}(v) = \frac{1 + \erf(v/\alpha)}{2} \xrightarrow{\quad \alpha\to0 \quad} H(v) = \begin{cases}1  &v>0, \\ 1/2  &v=0, \\ 0  &v<0, \end{cases}
\end{equation}  
that both converge (point-wise) to the Heaviside function in the limit $\alpha\to0$ (Fig.\,\ref{fig:sigmoids}). The parameter $\alpha$ represents the steepness of the sigmoid and the error function is defined as $\erf(z)=2/\sqrt{\pi}\int_{0}^{z}\exp{(-t^2)}dt$. Both sigmoids are well-defined as complex variable functions when $v\in\mathbb{R}$ is replaced by $z\in\mathbb{C}$ but, despite a similar real axis behavior and the same real axis limit, they behave quite differently in the complex plane, as shown in Fig.\;\ref{fig:erf_log}. Specifically, $\sigma_{\log,\alpha}(z)$ converges point-wise to a function that corresponds almost everywhere to $H(\mathrm{Re}(z))$,
\begin{equation}\label{eq:log_lim}
\lim_{\alpha\to0}\left[{\frac{1}{1+e^{-z/\alpha}}}\right] =
\begin{cases}
1 +i0 &x>0\\
0 +i0&x<0\\
\frac{1}{2}+i0 &x=0,y\neq-\pi\alpha -2\pi m\alpha\\
\infty + i 0 &x=0,y=-\pi\alpha -2\pi m\alpha.
\end{cases}
\end{equation}
On the other hand, observing the following decomposition of the error function $\erf(z/\alpha)$, 
\begin{equation}\label{eq:erf}
\erf(z/\alpha) = \frac{2}{\sqrt{\pi}}\int_{0}^{x/\alpha}e^{-t^2}\mathrm{d}t + \frac{2i}{\sqrt{\pi}}\int_{0}^{y/\alpha}e^{-(x/\alpha)^2+s^2}e^{-2ixs/\alpha}\mathrm{d}s,
\end{equation}
yields that in the region $\mathrm{Re}(z)^2-\mathrm{Im}(z)^2>0$, $\sigma_{\erf,\alpha}(z)$ converges to $1$ if $\mathrm{Re}(z)>0$ and to $0$ if $\mathrm{Re}(z)<0$, and outside the region $\mathrm{Re}(z)^2-\mathrm{Im}(z)^2>0$ no convergence is guaranteed.

By employing the sigmoid functions just introduced, we can then define 'smooth' cut-off distribution functions,
\begin{equation}
    f_{\erf,\alpha}(z) = \sigma_{\erf,\alpha}(z+v_c) \cdot F_0(z) \cdot \sigma_{\erf,\alpha}(-z+v_c), 
\end{equation}
and
\begin{equation}
    f_{\log,\alpha}(z) = \sigma_{\log,\alpha}(z+v_c) \cdot F_0(z) \cdot \sigma_{\log,\alpha}(-z+v_c), 
\end{equation}and observe the impact of different sigmoids on the solution of LVP.

Firstly, with the dominated convergence theorem, it can be proven that the LVP solution $E_{\alpha}(k,t)$ with $f_0=f_{\erf,\alpha}$ or $f_0=f_{\log,\alpha}$ converges to the same solution $E(k,t)$ with $f_0=f_{CO}$. Even if trivial in appearance, this fact ensures the good behavior of the LVP solution in the limit $\alpha\to0$, regardless of the chosen sigmoid.

Secondly,  as both $f_{\erf,\alpha}(z)$ and $f_{\log,\alpha}(z)$ fall into the entire/meromorphic functions class, the electric field solution can be entirely expressed as the sum of residues contributions, as per \eqref{eq:E_final}. However, because of the different behavior of the sigmoids in the complex plane, the root structure will look significantly different, depending on the sigmoid.
\begin{figure}
    \centering
    \begin{subfigure}{0.48\textwidth}
        \centering
        \includegraphics[width=\linewidth]{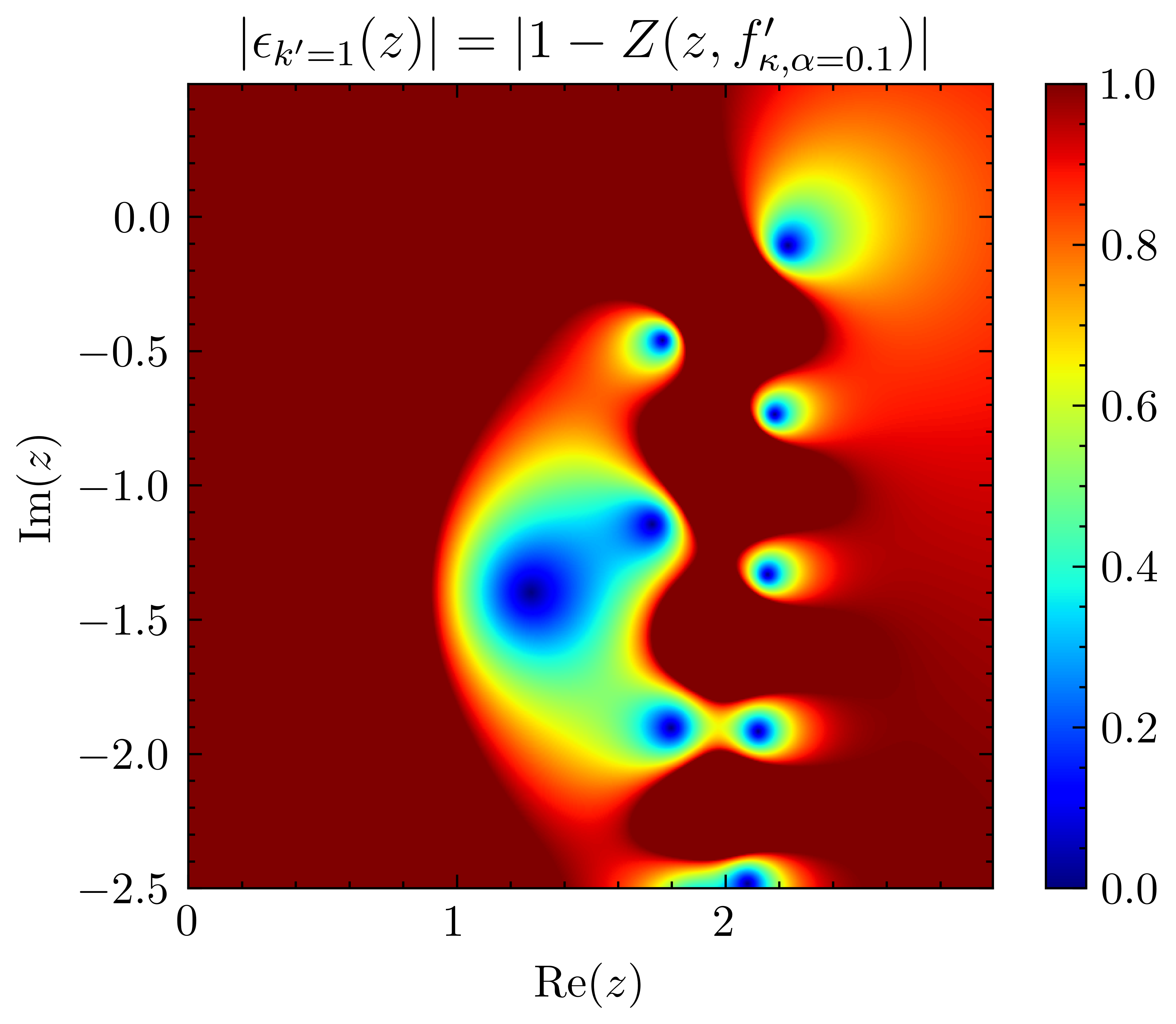} 
    \end{subfigure}
    \hfill
    \begin{subfigure}{0.48\textwidth}
        \centering
    \includegraphics[width=\linewidth]{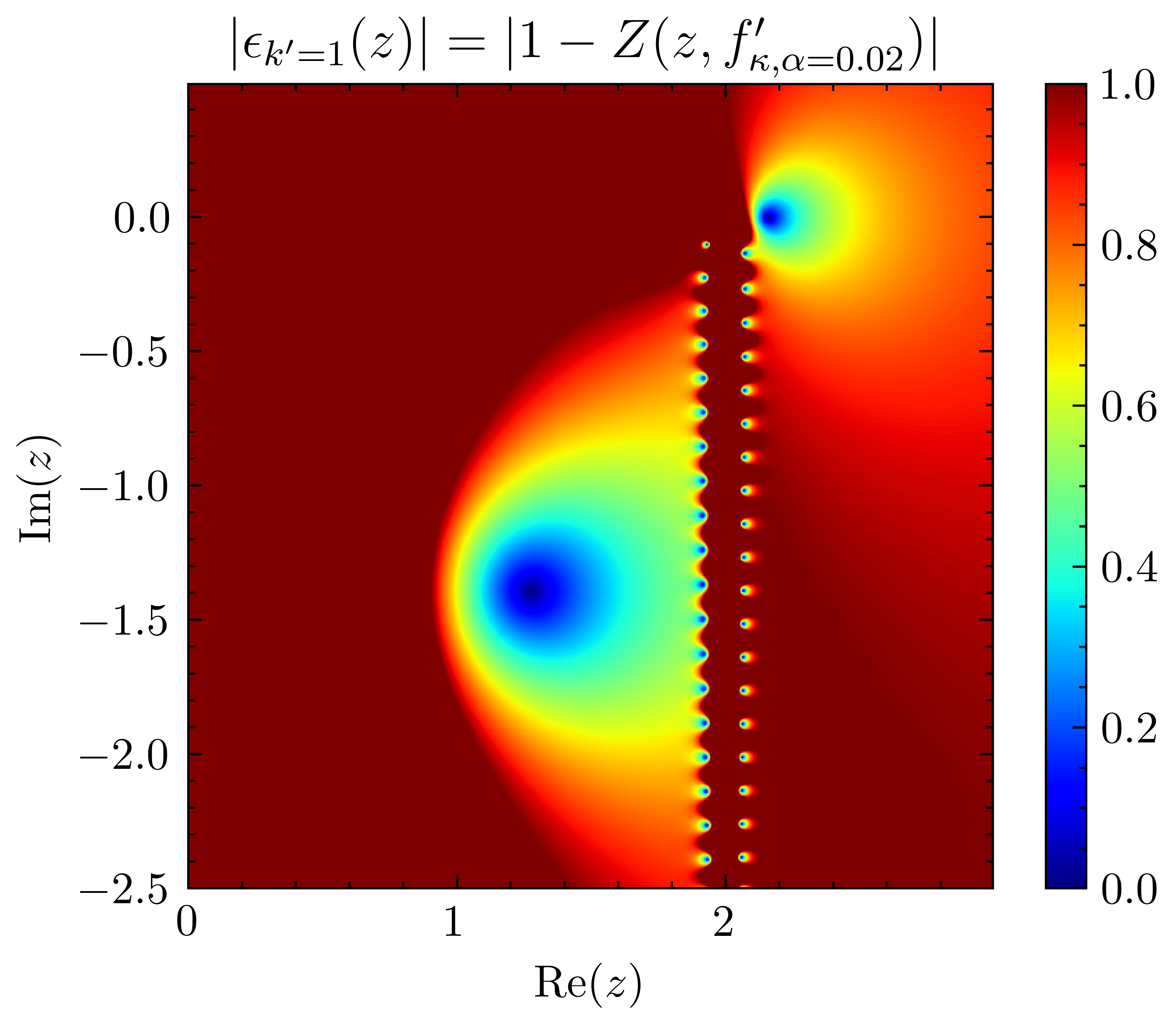} 
    \end{subfigure}
    \vspace{-0.5cm}
    \caption{Root structures for the smooth cut-off distribution $f_{log,\alpha}$ with $\alpha=0.1$ (\textbf{Left}) and $\alpha=0.02$ (\textbf{Right}). The support function $F_0$ is a $\kappa=1$ distribution with $v_t=\sqrt{2}$ and the cut-off is at $v_c=2$.}
    \label{fig:kappa_log}
\end{figure}

As a specific example, Fig.\;\ref{fig:kappa_log} shows the root structures for smooth cut-off distributions $f_{\log,\alpha}(z)$ for different $\alpha$ values, with the $\kappa=1$ distribution as support function $F_0$. By applying the concepts of Sec.\,\ref{Sec3}, the following comments can be made:
\begin{itemize}
    \item In both plots, we can recognize the left-most root as one of the two roots generated by the singularity of the support function derivative $F'_0=f'_{\kappa=1}$. 
    \item In both plots, the remaining roots are generated by the logistic sigmoid. According to \eqref{eq:log}, the singularities of $\sigma'_{\log}$ are located at $\mathrm{Re}{(z\pm v_c)}=0$ and $\mathrm{Im}(z)=-\alpha\pi(1+2m)$. Moreover, as can be easily seen by Taylor expanding $1/(1+e^{(-x/\alpha)})^2$, such singularities are of order 2 and generate 2 roots each.
    \item Among the sigmoid roots, the smallest damping one resembles the stable root of the $f_{CO,\kappa=1}$ case, shown in Fig.\,\ref{fig:kappa_co}.
    \item As $\alpha$ decreases, in agreement with the expression $\mathrm{Im}(z)=-\alpha\pi(1+2m)$, singularities and roots get denser and closer to $\mathrm{Re}(z)=v_c$. We reckon a similarity between the root structure of $f_{\log,\alpha}(z)$ in the limit $\alpha\to0$ and the root structure from $Z_1(f'_{CO})$, discontinuous at $\mathrm{Re}{(z)}=\pm v_c$ (center plot in Fig.\,\ref{fig:kappa_co}).
\end{itemize}

\begin{figure}
    \centering
    \begin{subfigure}{0.48\textwidth}
        \centering
        \includegraphics[width=\linewidth]{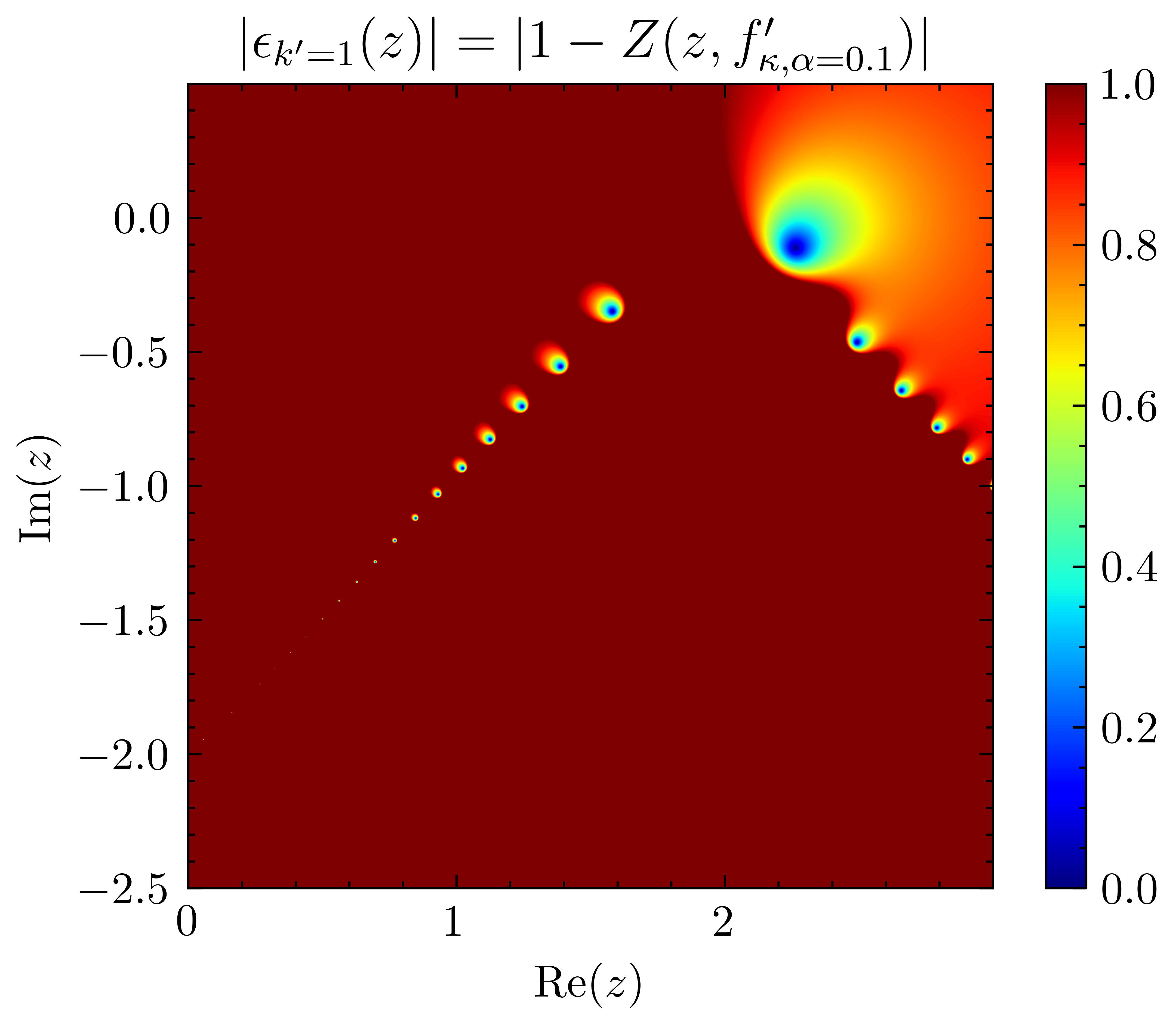} 
    \end{subfigure}
    \hfill
    \begin{subfigure}{0.48\textwidth}
        \centering
    \includegraphics[width=\linewidth]{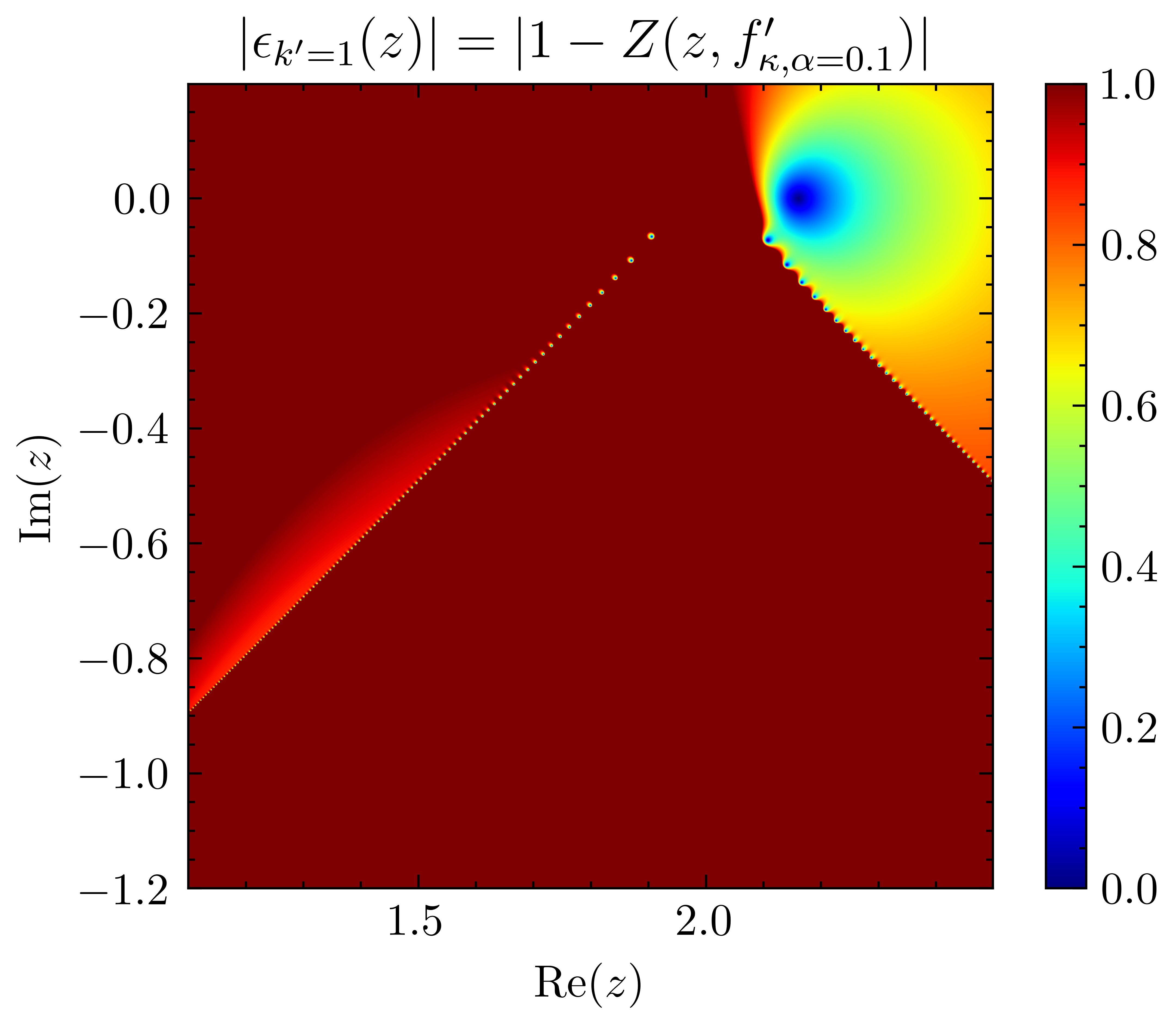} 
     \end{subfigure}
    \vspace{-0.5cm}
    \caption{Root structures for the smooth cut-off distribution $f_{\erf,\alpha}$,  $\alpha=0.1$ (\textbf{Left}) and $\alpha=0.02$ (\textbf{Right}). The support function $F_0(v)$ is a $\kappa=1$ distribution with $v_t=\sqrt{2}$ and the cut-off is at $v_c=2$.}
    \label{fig:kappa_erf}
\end{figure}

As a term of comparison, Fig.\,\ref{fig:kappa_erf} depicts the root structure of the same cut-off $\kappa=1$ distribution smoothed out by the error function sigmoid $\sigma_{\erf,\alpha}$, with the $\alpha$ parameters chosen so that $\sigma_{\erf,\alpha}(v)$ have qualitatively the same steepness of  $\sigma_{\log,\alpha}(v)$ used for Fig.\;\ref{fig:kappa_log}. The root structure is evidently dissimilar.
\begin{itemize}
\item The root generated by the singularity of $F'_0=f'_{\kappa=1}$ is no longer observable. 
\item The sigmoid function generates the remaining roots. In particular, as the derivative $\sigma_{\erf,\alpha}$ is proportional to $e^{-z^2}$, we notice a root structure similar to the Maxwellian case of Fig.\;\ref{fig:Max_Multipoles}. Moreover, the nearly stable root can still be spotted.
\item In the limit $\alpha\to0$, the sigmoid singularities become denser. However, they do not approach the same discontinuous structure as in the logistic case. Here, the sigmoid structure approaches a structure that is discontinuous in \textit{LHP} along the directions $\theta={-\pi/4,-3\pi/4}$, suggesting a new definition for the complex Heaviside function, i.e.
\begin{equation}
    H(z) = H(\mathrm{Im}(z)^2-\mathrm{Re}(z)^2) = \begin{cases}1  &\mathrm{Im}(z)^2>\mathrm{Re}(z)^2, \\ 0 &\mathrm{Im}(z)^2<\mathrm{Re}(z)^2. 
\end{cases}
\end{equation}

\end{itemize}

Based on the plots above, we are then led to conclude that the discontinuous dielectric functions associated with the Heaviside functions can be interpreted as the limiting case of 'sigmoid-smoothed' distribution functions. In particular, different sigmoid functions lead to different discontinuous dielectric functions, and the ambiguity in the complex definition of the Heaviside function can be connected to the multiple sigmoids one could use to replace $H(v)$. 
Moreover, we observe that by employing sigmoid functions, the discontinuity contribution $\Gamma_D$ from \eqref{eq:DR*} can be approximated and decomposed into the (infinite) sum of modes generated by the sigmoid function. As the sigmoid gets steeper, the infinite number of such solutions get denser and phase mix to yield the power law decay of the $\Gamma_D$ integral, previously shown by \cite{Hudson1962}.

\subsubsection{A privileged \textit{LHP} description?}
Although the two different sigmoid functions can be considered mathematically equivalent as they lead in the limit $\alpha\to0$ to the same electric field evolution, we wonder if one of them can be preferred. 
According to the common interpretation of Landau damping as a resonance effect, damping occurs because of the resonant interaction between the electric field component $\propto e^{\gamma t -i\omega t}$ and the particles with velocity $v\approx \omega/k$. Therefore, given a generic equilibrium distribution $f_0(v)=F_0(v)$ that features the set of roots $\mathcal{A}=\{p_n\}$, we expect the associated cut-off distribution $f_{CO}(v)=F_0(v) H(|v|<v_c)$ to yield the set $\mathcal{B}=\{p_n|v_c>|\mathrm{Re}{(p_n)}/k|\}$ in addition to the aforementioned stable solution. That is, the roots associated to the support function $F_0$ are roots for the cut-off distribution case if they resonate with the non-empty region of $f_{CO}$, i.e. $|\mathrm{Re}{(p_n)}/k|<v_c$. Consequently, the natural way of defining the complex Heaviside function according to the resonance interpretation would be $H(z)=H(\mathrm{Re}(z))$, as in the definition $W_{v_c,1}$.
When accounting for energy dispersion, the sigmoid $\sigma_{\log}$ is then preferred over $\sigma_{\erf}$ because of the limit 
$$
    \sigma_{\log,\alpha}(v)\to H(\mathrm{Re}(z)), \quad a.e.,
$$
and because the roots that $\sigma_{\log,\alpha}(z)$ introduces are all located in the region $\omega/k\approx v_c$, where the steep gradient acts as the sink of the wave damping process.

However, the resonance interpretation argument is flawed to the extent that the resonance explanation for Landau damping is usually based on the small damping assumption $|\gamma|\ll\omega$ while in the present case we are focusing on the behavior of heavily damped poles. Furthermore, the choice of the logistic sigmoid is not based on any a priori physical derivation and it was simply included in the discussions because of the relatively simple analytical expression among other sigmoid functions.  

On the contrary, the error function sigmoid appears analytically as one of the terms describing the energy dispersion for the fast-ions slowing-down distribution function from \cite{Gaffey1976}. It emerges by accounting for an additional term in the collisional operator of the fast-ion beam with the electron/ion background. While on one side, the similarity of the induced root structure to the Maxwellian one can be interpreted as the signature of collisions that smooth out the cut-off distribution function, on the other side, the existence of damped poles $\omega/k\gg v_c$ far from the cut-off is the source of contrast with the resonance interpretation. In fact, one wonders which particles  cause the damping since no such damped poles are present with the logistic sigmoid and  $\sigma_{\erf}(v)\ll\sigma_{\log}(v)$ for $|v|\gg v_c$.

Therefore, according to the present discussion, we cannot prefer one sigmoid over the other; further investigation is required.

\subsection{Slowing-down and non-integer $\kappa$ distributions}
\begin{figure}
    \centering
    \begin{subfigure}{0.48\textwidth}
        \centering
        \includegraphics[width=\linewidth]{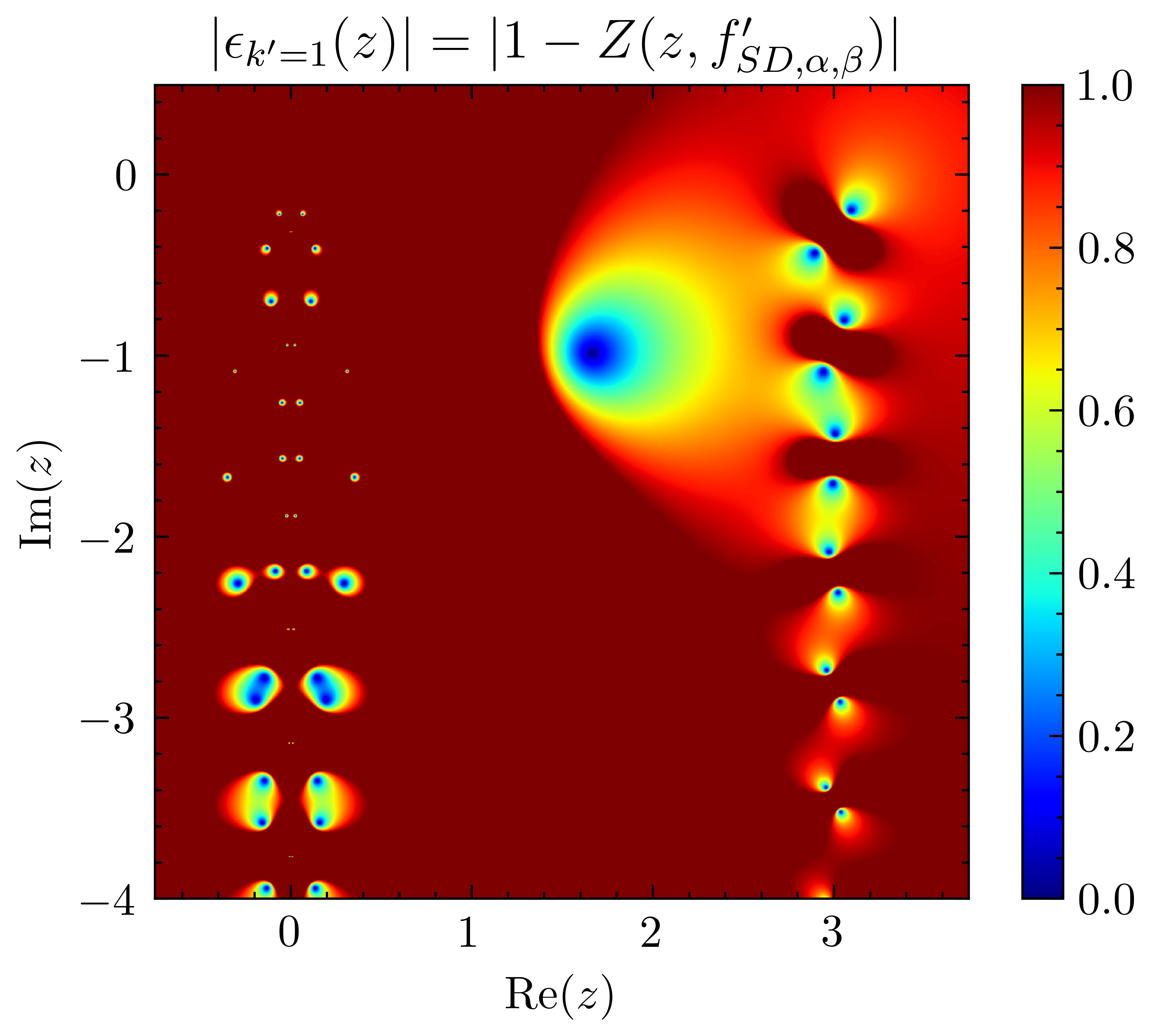} 
    \end{subfigure}
    \hfill
    \begin{subfigure}{0.51\textwidth}
        \centering
    \includegraphics[width=\linewidth]{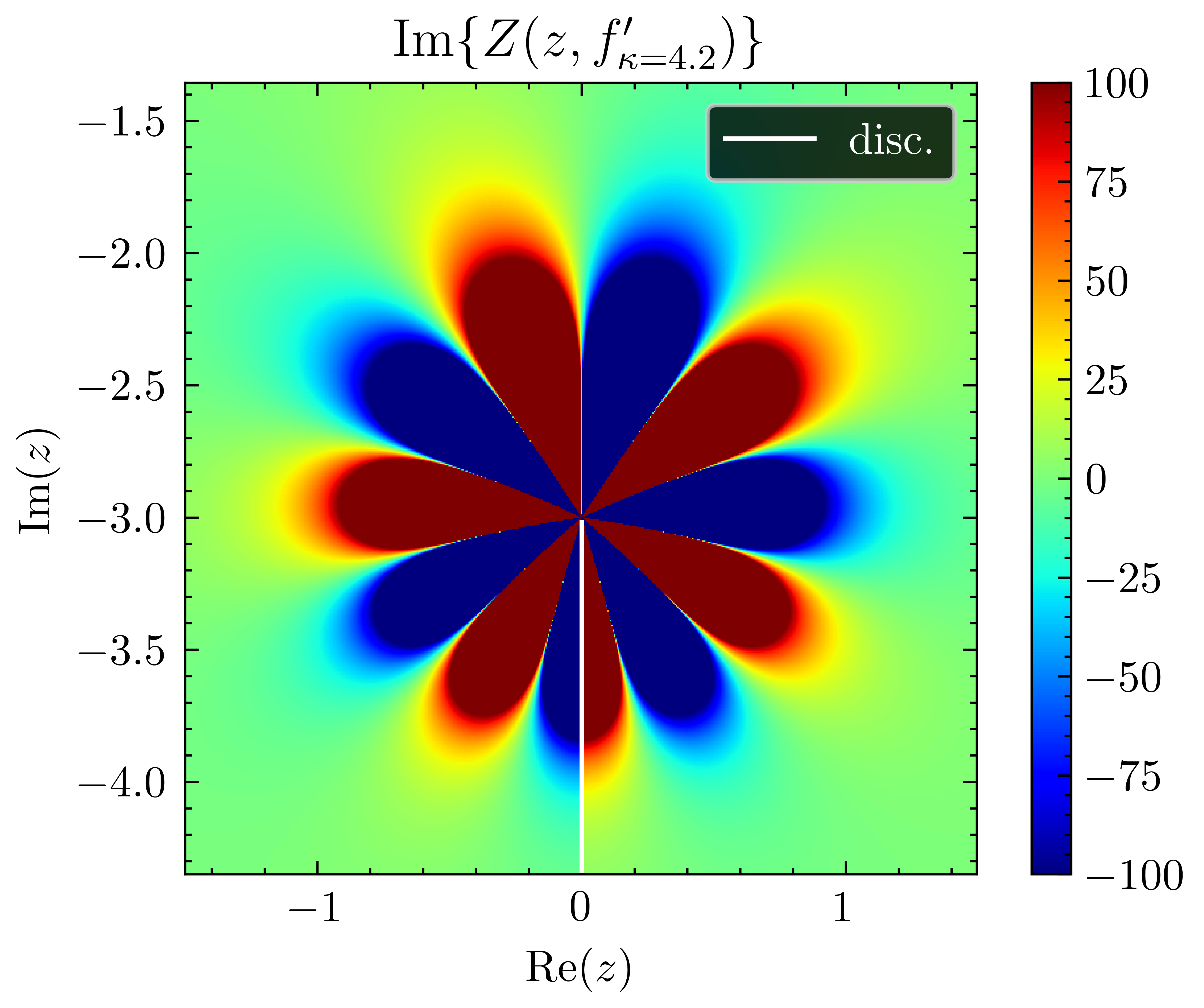} 
    \end{subfigure}
    \vspace{-0.5cm}
    \caption{(\textbf{Left}) Dispersion relation roots for the 'smooth' slowing down distribution. Logistic sigmoids are used, with steepness parameter $\alpha=\beta=0.1$. Other parameters: $v_t=\sqrt{2}$, $v_c=3$, $k'=1$. (\textbf{Right}) Imaginary part of the function $Z(z,f'_{\kappa=4.2})$. The white line represents the discontinuity.}
    \label{fig:SD_log}
\end{figure}

So far, we have focused on the non-analytical features of the LVP dispersion relation induced by the Heaviside function. However, other cases exist where the complex variable distribution features non-isolated points of discontinuities that are not determined by the Heaviside function. For example, the slowing-down distribution can be defined as in \cite{Xie2013},
\begin{equation}
    f_{SD}(v) = N(v_t,v_c) \frac{1}{|v|^3+v_t^3}W_{v_c}(v),
\end{equation}
where $v_t$ is a generic parameter, $v_c$ is the cut-off (injection) velocity, and $N$ is the normalization constant. Here, while the same discussions of the previous sections apply for $H_d$, the additional non-analytical feature is related to the modulus $|v|$. Without worrying about its meaning, the discontinuity induced by $|v|$ can be removed by noticing that
$$
    |v|=v\cdot\text{sign}(v)=v\cdot[2H(v)-1]
$$
and by simply approximating $H(v)$ with a sigmoid function. The left plot of Fig.\,\ref{fig:SD_log} shows the roots for $f_{SD,\alpha,\beta}$, corresponding to the 'smoothed' version of $f_{SD}$:
\begin{equation}
    f_{SD}(v) =  \sigma_{\alpha}(z+v_c) \cdot \frac{N(v_t,v_c)}{v^3\cdot[2\sigma_{\beta}(v)-1]^3+v_t^3} \cdot \sigma_{\alpha}(-z+v_c),
\end{equation}
with $\sigma_{\alpha}$ and $\sigma_{\beta}$ identifying two generic sigmoid functions with respective steepness parameter $\alpha$ and $\beta$. 
While the roots $\mathrm{Re}(z)\sim v_c$ roots relate to the $W_{v_c}$-induced discontinuity observed in the previous section, the $\mathrm{Re}(z)\sim 0$ roots relate to the $|v|$-induced discontinuity.

Another example worth mentioning corresponds to $\kappa$ distributions, defined in \eqref{eq:kappa_def}, with non-integer $\kappa$. The issue here arises because non-integer powers of complex numbers are multivalued and require the introduction of branch cuts.  Contrary to the discontinuities examined previously, we were not able to remove the non-integer $\kappa$ induced discontinuity (Fig.\,\ref{fig:SD_log}, right plot) and to decompose the $\Gamma_D$ integral into the sum of roots. Nevertheless, if we keep the same branch cut of Fig.\,\ref{fig:SD_log} running along the negative imaginary axis, it can be shown by simply plotting the absolute value of $\epsilon(z)$ for non-integer $\kappa$ distributions that the number of solutions in the small wavelength limit is $2\left(\left \lfloor{\frac{\kappa}{2}}\right \rfloor +1\right)$.

\section{Conclusions}
The present work focused on the full set of solutions of the linear Vlasov-Poisson system à la Landau. On a more practical level, we discussed how the multiple roots of the dispersion relation are generated, and for the case of entire/meromorphic functions we showed that the number of roots in the $k\to\infty$ limit is equal to the order of the singularity of the equilibrium distribution function derivative $f'_0(z)$. In the same limit $k\to\infty$ we also provided analytical formulas to describe roots for general meromorphic and Maxwellian distribution functions. On the other hand, for the case of the cut-off distribution function, we showed that the non-uniqueness of the associated lower half-plane description can be related to the different possible ways of describing the energy dispersion and that the sigmoid functions allow the decomposition of the $\Gamma_D$ integral as a more convenient sum of residues.

At a deeper level, the carried-out analysis made us ponder whether an improved understanding of the Landau damping phenomenon might hide behind the strongly damped roots. While remarking on the need for further and more rigorous investigation, we speculated about a possible connection between Landau damping and  the correlation among the degrees of freedom of the system, and we observed how the sigmoid-induced root structure questions the standard interpretation of Landau damping as a resonant mechanism. 

As a final comment, it is worth noticing that throughout the present work we assumed that the plasma under consideration could be described by a continuous (or piece-wise continuous) distribution function $f(v)$. However,  when $f(v)$ approaches zero, such as in the presence of steep sigmoid functions, this assumption might be debatable and an $N$-body discrete description on the model of \citet{Escande2018} might be more accurate. Therefore, we believe that analyzing the full structure of roots in terms of the $N$-body approach can be an interesting starting point to carry on the investigation on the open ends of the present work.

\section*{Acknowledgments}
Part of this work has been carried out within the framework of the EUROfusion 
Consortium, funded by the European Union via the Euratom Research and 
Training Programme (Grant Agreement No. 101052200—EUROfusion). Views and  
opinions expressed are however those of the author(s) only and do not 
necessarily reflect those of the European Union or the European 
Commission. Neither the European Union nor the European Commission can be
held responsible for them.

\bibliographystyle{abbrvnat}

\bibliography{biblio}

\begin{thebibliography}{14}
\providecommand{\natexlab}[1]{#1}
\providecommand{\url}[1]{\texttt{#1}}
\expandafter\ifx\csname urlstyle\endcsname\relax
  \providecommand{\doi}[1]{doi: #1}\else
  \providecommand{\doi}{doi: \begingroup \urlstyle{rm}\Url}\fi

\bibitem[Belmont et~al.(2008)Belmont, Mottez, Chust, and Hess]{belmont2008}
G.~Belmont, F.~Mottez, T.~Chust, and S.~Hess.
\newblock Existence of non-landau solutions for langmuir waves.
\newblock \emph{Physics of Plasmas}, 15\penalty0 (5):\penalty0 052310, 2008.

\bibitem[Escande et~al.(2018)Escande, B{\'e}nisti, Elskens, Zarzoso, and
  Doveil]{Escande2018}
D.~F. Escande, D.~B{\'e}nisti, Y.~Elskens, D.~Zarzoso, and F.~Doveil.
\newblock Basic microscopic plasma physics from n-body mechanics: A tribute to
  pierre-simon de laplace.
\newblock \emph{Reviews of Modern Plasma Physics}, 2:\penalty0 1--68, 2018.

\bibitem[Gaffey(1976)]{Gaffey1976}
J.~D. Gaffey.
\newblock Energetic ion distribution resulting from neutral beam injection in
  tokamaks.
\newblock \emph{Journal of Plasma Physics}, 16\penalty0 (2):\penalty0 149--169,
  1976.

\bibitem[Hudson(1962)]{Hudson1962}
J.~F. Hudson.
\newblock Landau damping for non-maxwellian distributions.
\newblock \emph{Mathematical Proceedings of the Cambridge Philosophical
  Society}, 58\penalty0 (1):\penalty0 119--129, 1962.

\bibitem[Landau(1946)]{Landau1946}
L.~Landau.
\newblock On the vibrations of the electronic plasma.
\newblock \emph{Journ. of Phys.}, 10\penalty0 (1):\penalty0 25--34, 1946.

\bibitem[Lazar and Fichtner(2021)]{Lazar2021}
M.~Lazar and H.~Fichtner.
\newblock \emph{Kappa Distributions}.
\newblock Springer, 2021.

\bibitem[Lima et~al.(2000)Lima, Silva~Jr, and Santos]{Lima2000}
J.~Lima, R.~Silva~Jr, and J.~Santos.
\newblock Plasma oscillations and nonextensive statistics.
\newblock \emph{Physical Review E}, 61\penalty0 (3):\penalty0 3260, 2000.

\bibitem[Livadiotis and McComas(2011)]{liv_2011}
G.~Livadiotis and D.~McComas.
\newblock Invariant kappa distribution in space plasmas out of equilibrium.
\newblock \emph{The Astrophysical Journal}, 741\penalty0 (2):\penalty0 88,
  2011.

\bibitem[Maekaku et~al.(2024)Maekaku, Sugama, and Watanabe]{maekaku2024time}
K.~Maekaku, H.~Sugama, and T.-H. Watanabe.
\newblock Time evolutions of information entropies in a one-dimensional
  vlasov--poisson system.
\newblock \emph{Physics of Plasmas}, 31\penalty0 (10), 2024.

\bibitem[Stix(1962)]{STIX}
T.~H. Stix.
\newblock \emph{The theory of plasma waves}.
\newblock McGraw-Hill (Advanced Physics Monograph Series), 1962.

\bibitem[Stucchi(2024)]{MyThesis2024}
R.~Stucchi.
\newblock Non-maxwellian distribution functions: Landau damping and beyond (ipp
  2024-17).
\newblock Master's thesis, Garching: Max-Planck-Institut für Plasmaphysik,
  July 2024.

\bibitem[Twiss(1952)]{Twiss1952}
R.~Twiss.
\newblock Propagation in electron-ion streams.
\newblock \emph{Physical Review}, 88\penalty0 (6):\penalty0 1392, 1952.

\bibitem[Weitzner(1963)]{Weitzner1963}
H.~Weitzner.
\newblock Plasma oscillations and landau damping.
\newblock \emph{The Physics of Fluids}, 6\penalty0 (8):\penalty0 1123--1127,
  1963.

\bibitem[Xie(2013)]{Xie2013}
H.-S. Xie.
\newblock Generalized plasma dispersion function: One-solve-all treatment,
  visualizations, and application to landau damping.
\newblock \emph{Physics of Plasmas}, 20\penalty0 (9):\penalty0 092125, 2013.

\end{thebibliography}

\end{document}